\documentclass[useAMS,usenatbib]{mn2e}
\usepackage{epsfig,subfigure,amsmath}
\usepackage{color}

\definecolor{purple}{rgb}{1,0,1}

\newcommand{\lcdm}{$\Lambda$CDM}
\newcommand{\hmpc}{$h^{-1}$Mpc}
\newcommand{\hgpc}{$h^{-1}$Gpc}
\newcommand{\hgpccubed}{$h^{-3}$Gpc$^{3}$}
\newcommand{\hmsol}{\mbox{ } {h}^{-1}~{M}_{\odot}}
\DeclareMathOperator\erf{erf}

\bibpunct[; ]{(}{)}{;}{a}{}{,}

\title[Voids in the SDSS DR9]{Voids in the SDSS DR9: observations, simulations, and the impact of the survey mask}
\author[P.~M. Sutter et al.]
{
\parbox{\textwidth}{
{P.~M. Sutter}$^{1,2,3,4}$ \thanks{Email: sutter@iap.fr},
Guilhem Lavaux$^{1,2,5,6,7}$,
Benjamin D. Wandelt$^{1,2,4,8}$,
David H. Weinberg$^{3,9}$, 
Michael S. Warren$^{10}$, and
Alice Pisani$^{1,2}$
}
\vspace{0.4cm}\\
\parbox[c]{\textwidth}{
$^{1}$ Sorbonne Universit\'{e}s, UPMC Univ Paris 06, UMR7095, Institut d'Astrophysique de Paris, F-75014, Paris, France \\
$^{2}$ CNRS, UMR7095, Institut d'Astrophysique de Paris, F-75014, Paris, France \\
$^{3}$ Center for Cosmology and Astro-Particle Physics, Ohio State University, Columbus, OH 43210\\
$^{4}$ Department of Physics, University of Illinois at Urbana-Champaign, Urbana, IL 61801\\
$^{5}$ Department of Physics \& Astronomy, University of Waterloo, Waterloo,
ON,  N2L 3G1 Canada \\
$^{6}$ Perimeter Institute for Theoretical Physics,
Waterloo, ON, N2L 2Y5, Canada \\
$^{7}$ Canadian Institute for Theoretical Astrophysics, 60 St. George St.,
Toronto, ON M5S 3H8 Canada \\
$^{8}$ Department of Astronomy, University of Illinois at Urbana-Champaign, Urbana, IL 61801\\
$^{9}$ Department of Astronomy, Ohio State University, Columbus, OH 43210\\
$^{10}$ Theoretical Division, Los Alamos National Laboratory, Los Alamos, NM 87545, USA
}}

\begin{document}

\maketitle

\label{firstpage}

\begin{abstract}
We present and study cosmic voids identified using the watershed 
void finder {\tt VIDE} in the Sloan Digital Sky 
Survey Data Release 9, compare these voids to ones identified 
in mock catalogs,
and assess the impact of the survey mask on void 
statistics such as number functions, ellipticity distributions, 
and radial density profiles. 
The nearly 1,000 identified voids span 
three nearly volume-limited samples from redshift $z=0.43$ to $0.7$.
For comparison we use 98 of the 
publicly available $2LPT$-based mock galaxy catalogs of 
Manera et al., and also generate our own mock catalogs 
by applying a Halo Occupation Distribution model to an $N$-body 
simulation. 
We find that the mask reduces the number density of voids at all scales 
by a factor of three and
slightly skews the relative size distributions. This
engenders an 
increase in the mean ellipticity by roughly $30\%$.  
However, we find that radial density profiles are 
largely robust to the effects of the mask.
We see excellent agreement between the data and both mock catalogs, 
and find no tension between the observed void properties and the properties 
derived from $\Lambda$CDM simulations.
We have added the void catalogs from both data and mock 
galaxy populations discussed in this work 
to the Public Cosmic Void Catalog at http://www.cosmicvoids.net.
\end{abstract}


\section{Introduction}

With the recent advent of large-scale comprehensive void 
catalogs~\citep{Pan2011,Sutter2012a}, 
cosmological analysis is beginning to fan out from probes solely 
focused on overdensities such as galaxy correlations~\citep{Sanchez2012,Marin2013}
 and baryon acoustic oscillations~\citep{Bassett2010}
 to more general studies based on 
alternative information sources available in the cosmic web.
Since the primary target of cosmological analysis is often quantifying 
and understanding dark energy~\citep{Weinberg2012}, 
exploiting the underdense voids
in the matter distribution of the universe is a natural choice:
the interiors of voids are dominated by dark energy~\citep{Goldberg2004}, 
so their shapes, sizes, and growth histories are 
intimately tied to the global properties of the large-scale 
universe~\citep{Thompson2011}.

Already researchers have begun to exploit the public void 
catalogs. 
\citet{Ilic2013} correlated void positions with 
WMAP measurements of 
Cosmic Microwave Background temperature anisotropies~\citep{Komatsu2011}
to obtain 
a weak measurement of the integrated Sachs-Wolfe 
effect~\citep{Thompson1987}. 
The Planck Collaboration followed up on this study to confirm a 
detection~\citep{Planck2013b}.
\citet{Melchior} have performed a measurement of 
gravitational weak lensing around voids in data
(theoretically predicted by~\citealt{Krause2013} and 
~\citealt{Higuchi2013}) 
to directly measure the underdensities in the 
dark matter.
~\citet{Pisani2013} 
used a novel method to measure the real-space radial density 
profiles within voids without assumptions about 
cosmological or redshift-space distortion models.
Finally,~\citet{Sutter2012b} began to measure cosmological 
parameters by leveraging the statistical isotropy of stacked 
voids~\citep{LavauxGuilhem2011} 
to perform an Alcock-Paczynski test~\citep{Alcock1979, Ryden1995}.

Looking ahead, there are many more promising applications of voids 
to cosmology and astrophysics. 
At the most simple level, the size distribution of voids is sensitive
to cosmological parameters~\citep{Jennings2013} 
and modified gravity~\citep{Clampitt2013}, though 
these effects can be confused by uncertainties in galaxy formation 
physics~\citep{Little1994, Muller2000, Tinker2009}.
A measurement of the shapes 
of voids as encoded by the mean ellipticity would shed light 
on dark energy~\citep{Biswas2010,Bos2012}
as well as the two-point correlation of the void 
positions~\citep{Padilla2005,Paranjape2012,Hamaus2013}. 
The radial density profile, reconstructed in real space using techniques 
such as those described in~\citet{Pisani2013}, can also be used to constrain 
exotic dark energy models~\citep{Shoji,Spolyar2013}.
Astrophysically, voids can also be used to measure primordial 
magnetic fields~\citep{Taylor2011,Beck2013}
and the effects of environment on galaxy 
formation~\citep{Gottlober2003,Rojas2004,Hoyle2005,Rojas2005,VandeWeygaertR.2011,Ceccarelli2012,Hoyle2012}

To support current and future void-based science efforts we must 
continue to identify voids in the latest galaxy surveys.
This way we can take advantage of deeper and wider surveys for a
greater redshift lever arm for cosmological parameter estimation
and for more volume for increasing the signal-to-noise of statistical void 
properties.
Also, even though current surveys such as the Baryon Oscillation
Spectroscopic
Survey (BOSS;~\citealt{Dawson2013}) may not be optimal for void 
analysis due to their relatively low sampling density, 
we can use void catalogs from data to test and calibrate results 
against theoretical expectations in preparation for larger-volume
surveys in the future such as Euclid~\citep{Laureijs2011}, 
BigBOSS~\citep{Schlegel2011}, and 
WFIRST~\citep{Spergel2013}.

Ever since~\citet{Peebles2001} pointed out a potential discrepancy 
between the interior contents of voids in \lcdm~predictions and 
observations, there has been intense interest in comparing voids 
between simulations and observations.
This has been done for earlier void catalogs in 
the 2-Degree Field Galaxy Redshift 
Survey (2dFGRS;~\citealt{Benson2003,Hoyle2004,Ceccarelli2006})
and the Sloan Digital Sky Survey (SDSS;~\citealt{Strauss2002}) 
Data Release 7~\citep{Pan2011}, 
and in earlier surveys~\citep{Einasto1991,Weinberg1992,Little1994,Vogeley1994}
but in a very restricted context:
it is difficult to build simulations with high enough resolution to 
capture all the survey galaxies and sufficient size to enclose the 
entire survey volume. Rather than attempt to reproduce complex 
observational details such as survey geometry, 
typically authors take a limited volume 
within the survey and compare the statistical properties of the 
remaining voids to voids identified in a galaxy population generated 
with semi-analytic modeling~\citep[e.g.,][]{Tavasoli2013}.
This common approach has several shortcomings: 
it is difficult to precisely tune semi-analytic 
models to a given survey~\citep{Baugh2003}
and it does not take advantage of the full survey volume.
We can address any potential discrepancies in a more robust way 
by building nearly identical survey-like populations in 
our simulations.

An examination of the impacts of the survey mask is especially important, 
since only selecting voids far away from the survey area discards 
much useful information, and without rigorous void selection 
there may still be residual systematics. Also, since 
theoretical work with voids is done in simulations with cubic 
volumes, understanding 
the role of the mask is essential for building the links between 
theory and data. Since survey masks usually have complicated shapes, 
their impact is highly non-trivial, non-obvious, and different for each 
survey. 
\citet{vonBenda2007} noted differences between masked and unmasked 
void populations in the 2dFGRS,
although~\citet{Pan2011} did not find significant 
differences when examining the properties of voids with their void 
finding algorithm in a low-redshift volume-limited sample of SDSS 
galaxies.
However, there has been no such examination 
in higher redshifts of the SDSS with the {\tt VIDE} 
algorithm~\citep{Sutter2014c}, 
which is the source of the current 
large void catalogs.

We explore another important link, the impacts of 
sparsity and galaxy bias, in another work~\citep{Sutter2013a}, 
while earlier works 
such as~\citet{Ryden1996} 
have connected redshift-space voids to those in real 
space. 

In this work, we present voids in the SDSS
Data Release 9 CMASS sample~\citep{Ahn2012}, 
a survey covering 3,000 square degrees from 
redshift 0.43 to 0.7, for a total volume of nearly $1.5$ cubic \hgpc. 
We compare these voids to voids found in two sets of mock catalogs: 
the published mocks of~\citet{Manera2013} and our own derived from a
large-volume high-resolution $N$-body simulation. In the simulation 
we are able to capture the entire survey without overlapping 
or stitching simulation volumes, allowing us to examine 
the systematic impacts of the survey mask on void statistical 
properties such as number functions, radial profiles, and 
ellipticities.
While the low galaxy density of this survey is not ideal for 
void identification, the analysis of~\citet{Sutter2013b} indicates that 
voids found in surveys of this type still correspond to 
physical underdensities in the dark matter, and thus are still 
useful probes of cosmology and astrophysics.
In addition,~\citet{Sutter2013a} shows that the universal density 
profile of~\citet{Hamaus2014} fits voids identified in all densities
of samples, and that there exist simple scaling relations 
between voids in different samples, which means that the objects 
identified even in low-density surveys correspond to voids.

In the next Section we establish our coordinate system and briefly 
discuss our void-finding method and strategies for handling 
masks in the survey data. Section~\ref{sec:voiddata} introduces
our galaxy survey samples and the properties of the voids 
identified in them.
In Section~\ref{sec:voidmocks} 
we present our mock galaxy populations
and compare voids in these masked and unmasked populations 
to the voids in the data.
Finally, Section~\ref{sec:conclusions} offers concluding remarks regarding
implications for future surveys and void-based science.

\section{Void finding}
\label{sec:voidfinding}

For each galaxy in the survey, we transform its sky latitude
$\theta$, sky longitude $\phi$, and 
redshift $z$,  to a comoving coordinate system:
\begin{eqnarray*}
  x' & = & D_c(z) \cos{\phi} \cos{\theta}, \\
  y' & = & D_c(z) \sin{\phi} \cos{\theta}, \\
  z' & = & D_c(z) \sin{\theta}, 
\label{eq:transform}
\end{eqnarray*}
where $D_c(z)$ is the comoving distance to the galaxy 
at redshift $z$.
We assume a \lcdm~cosmology consistent with WMAP 7-year
results~\citep{Komatsu2011}: $\Omega_M=0.27$, $\Omega_\Lambda=0.73$,
and $h=0.71$. 

We identify voids with a modified and extended version of 
{\tt ZOBOV}~\citep{Neyrinck2008, LavauxGuilhem2011, Sutter2012a}.
called {\tt VIDE}~\citep{Sutter2014c}.
{\tt VIDE} creates a Voronoi tessellation of the tracer particle 
population and uses the watershed transform to group Voronoi 
cells into zones and voids~\citep{Platen2007}. 
The watershed transform identifies catchment basins
as the cores of voids and ridgelines, which separate the flow
of water, as the boundaries of voids.
The watershed transform builds a nested hierarchy of
voids~\citep{LavauxGuilhem2011, Bos2012}, and for the purposes
of this work we only examine \emph{root} voids, which
are voids at the base of the tree hierarchy and hence have no parents.
We also impose two density-based criteria on our void catalog.
The first is a threshold cut within {\tt VIDE} itself where
voids only include as additional members Voronoi zones with density
less than $0.2$ the mean particle density.
If a void consists of only a single zone (as they often do in 
sparse populations) then this restriction does not apply.
We apply the second density criterion as a post-processing step:
we only include voids with mean central densities below $0.2$
the mean particle density. We measure this central density
within a sphere with radius $R = 0.25 R_{\rm eff}$, where
\begin{equation}
  R_{\rm eff} \equiv \left( \frac{3}{4 \pi} V \right)^{1/3}.
\label{eq:reff}
\end {equation}
In the expression above, $V$ is the sum of the Voronoi volumes of the 
particles which comprise the void.
We also ignore voids with $R_{\rm eff}$ below the mean particle
spacing of the tracer population.

Additionally, for the analysis below we need to define a center for the
void. For our work, we take the macrocenter, or volume-weighted
center of all the Voronoi cells in the void:
\begin{equation}
  {\bf X}_v = \frac{1}{\sum_i V_i} \sum_i {\bf x}_i V_i,
\label{eq:macrocenter}
\end{equation}
where ${\bf x}_i$ and $V_i$ are the positions and Voronoi volumes of
each tracer $i$, respectively.

\begin{figure} 
  \centering 
  {\includegraphics[type=png,ext=.png,read=.png,width=\columnwidth]{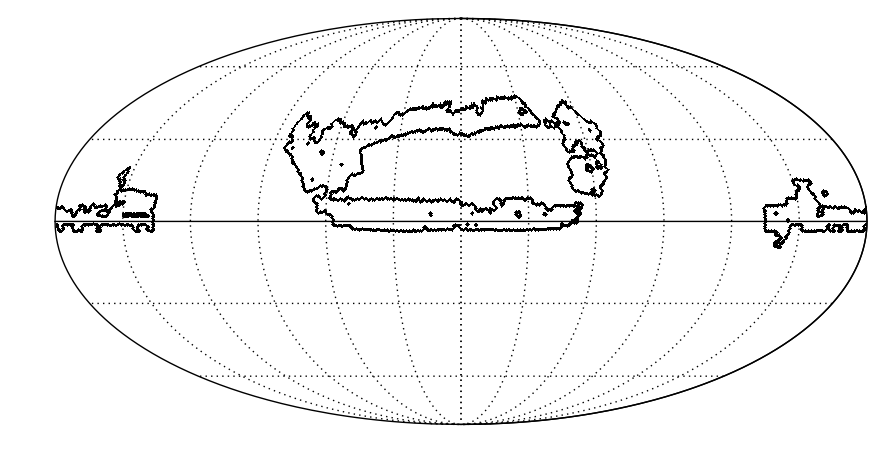}}
  \caption{ {\tt HEALPix} map in a Mollweide projection
                  of identified boundary zones (black) around and
                  within the SDSS DR9 survey area where 
                  we inject boundary particles.
          }
\label{fig:mask}
\end{figure}

As presented in~\citet{Sutter2012a} and~\citet{Sutter2014c}, 
{\tt VIDE} includes modifications to {\tt ZOBOV} to 
account for survey boundaries, internal masks, and redshift limits.
To handle line-of-sight boundaries and internal masks, we pixelize the survey 
region 
using {\tt HEALPix}~\citep{Gorski2005}\footnote{http://healpix.jpl.nasa.gov}
and identify boundary pixels (i.e., pixels with at least one 
non-survey region neighbor).
We inject particles along the line of sight within each 
boundary pixel with a spatial 
density of $10^{-3} (h^{-1} \rm {Mpc})^{-3}$.
By giving these boundary particles essentially infinite density and 
breaking their degeneracies in the Voronoi graph, 
we prevent the watershed algorithm from growing voids outside the 
survey region. 
Also, these boundary particles serve as a marker for identifying 
voids near the edge.
Figure~\ref{fig:mask} shows our identification of the SDSS DR9~\citep{Ahn2012}
survey boundary pixels.
To accurately capture
the shape of the mask we required a resolution of $N_{\rm side}=512$ 
($\sim 2-3$~\hmpc~at $z=0.7$ in a \lcdm~universe).

Following the procedure of~\citet{Sutter2012a},
we generate two void catalogs for each sample, which 
we call \emph{all} and ~\emph{central} catalogs.
Naturally, the \emph{all} sample contains every identified void 
which satisfied the density cutoff criteria, even voids near the 
survey edge. On the other hand, \emph{central} voids do not 
touch any boundary particle (i.e., the most distant void member particle is 
closer than any boundary particle) , and thus are not near any survey 
boundary or internal mask. 
The \emph{central} sample is designed to ensure 
that we have a fair distribution of void shapes and
alignments within the survey volume.
To handle survey high-redshift caps, we simply exclude from all catalogs 
any void which extends beyond the redshift limits of a
 given sample. 
This is a more restrictive approach than the procedure discussed 
in~\citet{Sutter2012a}, since it is difficult to construct stable 
tessellations on co-spherical points.
To evaluate this, 
if the distance from the void macrocenter to the 
redshift edge is closer than the distance to \emph{any} void member 
particle, 
we reject the void.

\section{Voids in Data}
\label{sec:voiddata}

\subsection{Galaxy Populations}
We take our galaxy sample from the CMASS selection of SDSS DR9 (BOSS)
spectroscopic targets~\citep{Ahn2012}. This is the same sample of 
galaxies used in the analysis of~\citet{Reid2012}.
The 455,281 galaxies in this selection of the survey 
extend from $z=0.43$ to $z=0.7$.
As before in our catalog of voids in SDSS DR7~\citep{Sutter2012a}, 
we take volume-limited samples to ensure statistical uniformity
and constant shot noise and galaxy bias around our identified voids.
Also, we require volume-limited samples in order to compare to
our mocks based on Halo Occupation Distribution (HOD) models, 
which are constructed around volume-limited surveys~\citep{Berlind2002}.

We apply simple evolution and $K$-corrections of the form
\begin{eqnarray}
  z_e & = & -2 (1-z+0.1)(z-0.1)\\
  z_k & = & -0.242659 + 1.38731 z,
\end{eqnarray}
and
compute absolute magnitudes $M_r$ assuming
the cosmological parameters noted above.
We choose three redshift bins.
Each redshift range is characterized by a typical galaxy luminosity,
which we differentiate by the labels~\emph{dim}, \emph{mid}, 
 and~\emph{bright}.
Our redshift bins are:
$0.43 < z < 0.5$, which we label \emph{CMASS Dim},
$0.5 < z < 0.6$, labeled as \emph{CMASS Mid},
and $0.6 < z < 0.7$, called \emph{CMASS Bright}.
Table~\ref{tab:samples}
lists the sample name, limiting absolute magnitude,
redshift bound, comoving volume, number of galaxies, and
 the mean galaxy separation in that sample.
The mean galaxy spacing is $(n_g/V)^{-1/3}$, where
$n_g$ is the number of galaxies within each sample and $V$ is the
sample volume.
\begin{table*}
\centering
\caption{Volume-limited galaxy samples used in this work.}
\tabcolsep=0.11cm
\footnotesize
\begin{tabular}{ccccccc}
  Sample Name & $M_{r, {\rm max}}$ & $z_{\rm min}$ & 
              $z_{\rm max}$ & Volume & $N_{\rm gal}$ &  $\bar{n}^{-1/3}$ \\
   &  &  & & (\hgpccubed) &  & (\hmpc) \\
  \hline  \hline
CMASS Dim & -20.1 & 0.43 & 0.5 & 0.29 &  61249 & 16.76\\
CMASS Mid & -20.1 & 0.43 & 0.6 & 0.82 & 188300 & 16.29\\
CMASS Bright & -20.8 & 0.43 & 0.7 & 1.48 & 205732 & 19.29\\
\hline
\end{tabular}
\label{tab:samples}
\end{table*}

\begin{figure} 
  \centering 
  {\includegraphics[type=png,ext=.png,read=.png,width=\columnwidth]{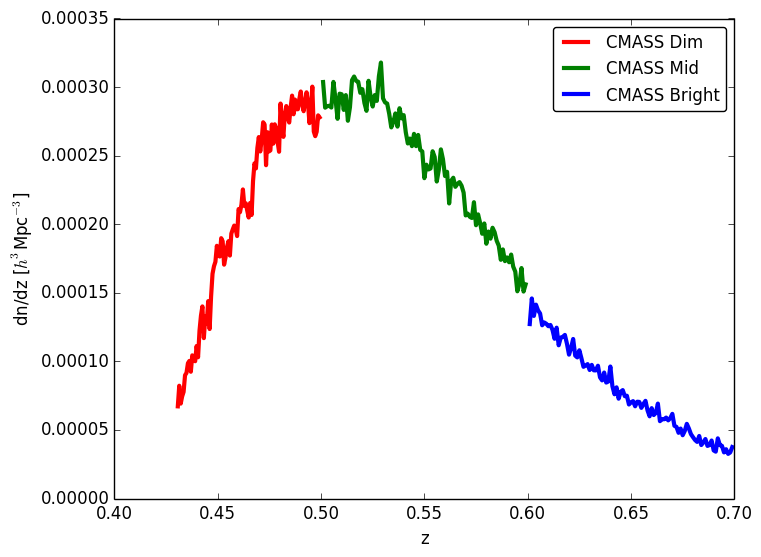}}
  \caption{ Density of galaxies as a function of redshift in our three 
            nearly volume-limited samples.
          }
\label{fig:galden}
\end{figure}

Figure~\ref{fig:galden} shows the galaxy number density as a function of 
redshift for each of our samples. These plots show that a simple 
luminosity cut does not produce a truly volume-limited sample due to the 
complex target selection procedure in CMASS. The density in the \emph{Dim} 
sample varies by a factor of $\sim 4$, in the \emph{Mid} sample by a factor 
of $\sim 2$, and in the \emph{Bright} sample by a factor of $\sim 3$.
While we can account for redshift dependence in the densities by 
weighting individual galaxies in the {\tt VIDE} code~\citep{Neyrinck2008}, 
our testing has shown 
that this does not strongly impact void properties, for two reasons.
First, only the largest voids will span a deep enough redshift range 
to be sensitive to changes in the underlying density, and since the number 
of large voids are exponentially suppressed (and they are more likely to be 
removed from consideration because they intersect an internal boundary), varying galaxy number 
density will only affect a small percentage of voids. 
Secondly, the nature of the watershed algorithm naturally guards against  
effects of varying density: since each particle has on average 17
adjacent particles~\citep{Neyrinck2008}, we must reduce the local 
density of a wall by a large factor before it is no longer identified 
as a void boundary.

Table~\ref{tab:voidsamples_data} summarizes the data samples used in this work,
the redshift ranges used to produce the void samples, 
their respective volumes, and the total number of voids identified 
in each sample. In total, we identify nearly $1,000$ voids in the 
\emph{all} sample, while the \emph{central} sample produced
$\sim 480$ voids.

\begin{table}
\centering
\caption{Summary of voids in data.}
\tabcolsep=0.11cm
\footnotesize
\begin{tabular}{cccccc}
  Sample Name & $z_{\rm min}$ & $z_{\rm max}$ & Volume & $N_{\rm voids}$ &\\
   &  &  & (\hgpccubed) &  &\\
  \hline  \hline
CMASS Dim, all & 0.43 & 0.5 & 0.29 & 283 &  \\ 
CMASS Dim, central & 0.43 & 0.5 & 0.29 & 151  & \\ 
CMASS Mid, all & 0.5 & 0.6 & 0.53 & 570 &  \\ 
CMASS Mid, central & 0.5 & 0.6 & 0.53 & 242  & \\ 
CMASS Bright, all & 0.6 & 0.7 & 0.66 & 283  & \\ 
CMASS Bright, central & 0.6 & 0.7 & 0.66 & 137  & \\ 
\hline
\end{tabular}
\label{tab:voidsamples_data}
\end{table}

\subsection{Void Properties}

\begin{figure} 
  \centering 
  {\includegraphics[type=png,ext=.png,read=.png,width=\columnwidth]{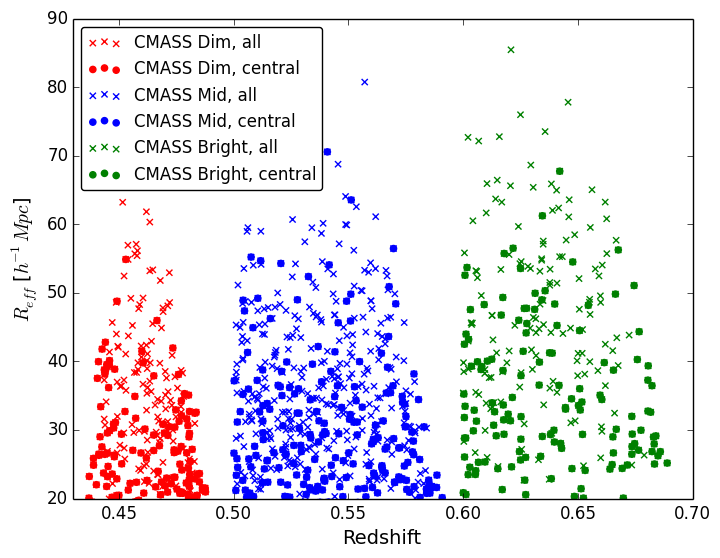}}
  \caption{Distribution of voids within the CMASS samples. We plot 
           void effective radius $R_{\rm eff}$ (Eq.~\ref{eq:reff})
           versus redshift. We show the voids in the volume-limited 
           samples \emph{CMASS Dim} (red), \emph{CMASS Mid} (blue), and
           \emph{CMASS Bright} (green). All voids are marked with 
           a cross, and voids in the \emph{central} catalog are 
           marked by filled circles.
          }
\label{fig:rvsz}
\end{figure}

Figure~\ref{fig:rvsz} shows the distribution of void sizes as a 
function of redshift for all the galaxy samples. We show 
both \emph{all} and \emph{central} voids. We see that though a few 
voids in \emph{CMASS Mid} and \emph{CMASS Bright} reach an effective 
radius of $\sim 80$ \hmpc, most --- especially in the \emph{central} 
catalog --- are below 50 \hmpc. For \emph{Mid} and \emph{Bright} 
samples, voids pervade the 
low-redshift boundary, because here we keep the population 
of galaxies below that boundary and only reject voids whose 
centers fall below the redshift cutoff. On the other hand, we see 
a tapering in the distribution at the high-redshift caps:
here, we reject any void that might intersect the cap, and 
we are more likely to cut progressively larger voids. 
We remove even more of the largest voids when creating the 
\emph{central} samples, since these voids are more likely 
to lie nearby the mask line-of-sight boundaries.
We observe a distinct lack of small voids in the \emph{Bright}
sample, which is a consequence of its slightly lower 
mean galaxy spacing. While we expect to see smaller voids 
at higher redshift, the effects of sparsity and biasing lead to 
larger observed voids~\citep{Daloisio2007}.

\begin{figure} 
  \centering 
  {\includegraphics[type=png,ext=.png,read=.png,width=\columnwidth]{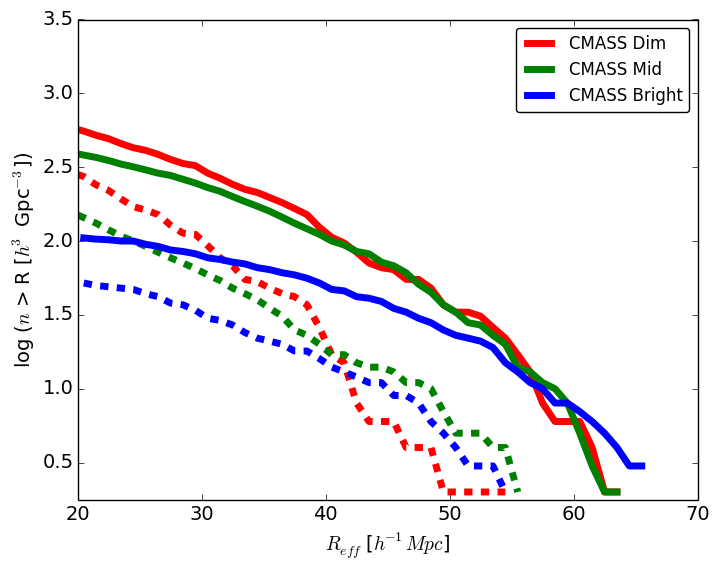}}
  \caption{Cumulative void number function for \emph{CMASS Dim} (red),
           \emph{CMASS Mid} (blue), and \emph{CMASS Bright} (green).
           The number functions of all voids, including those near survey 
            boundaries and internal masks, are shown as solid lines, 
           while \emph{central} catalog voids are shown as dotted lines.
          }
\label{fig:numberfunc_data}
\end{figure}

We show another way of expressing the size distribution in
Figure~\ref{fig:numberfunc_data}. 
This is a plot of the cumulative 
number function: the total number of voids in each sample above 
a given effective radius. The number function is a potentially 
powerful probe of cosmology with voids~\citep{Sheth2004}.
All samples have roughly the same number density
of the largest voids, but the lack of smaller voids in the \emph{CMASS Bright}
sample manifests as a uniformly reduced number function for small- and 
medium-scale voids. In all samples, the \emph{central} catalogs host 
roughly half as many voids as the \emph{all} catalogs. 
This is a consequence of the relatively large surface-to-volume 
ratio of the current CMASS survey. 
The relatively narrow angular extent especially impacts our 
void populations; the small internal holes do little disruption.
As future data releases fill 
in the central regions of the expected coverage area, we 
should approach the higher fraction of \emph{central} voids 
seen in earlier surveys~\citep{Sutter2012a}.

Though voids have complex shapes, we can assign them a 
unique ellipticity. This simple scalar captures most of the 
shape information of the void, and its distribution is a 
sensitive cosmological probe~\citep{Park2007,Biswas2010,Lavaux2010,Bos2012}.
To compute the ellipticity, for a given
set of galaxies within a void we first construct the 
inertia tensor:
\begin{eqnarray}
  M_{xx} & = &\sum_{i=1}^{N_p} (y_i^2 + z_i^2) \\ 
  M_{xy} & = & - \sum_{i=1}^{N_p} x_i y_i, \nonumber
\end{eqnarray}
where $N_p$ is the number of galaxies in the void, and
$x_i$, $y_i$, and $z_i$ are coordinates of the particle $i$
relative to the void center.
The other components of the tensor are obtained by 
cyclic permutations. 
Given the inertia tensor, we compute the eigenvalues and form
the ellipticity:
\begin{equation}
  \epsilon = 1- \left( \frac{J_1}{J_3}\right)^{1/4}, 
\label{eq:ellip}
\end{equation}
where $J_1$ and $J_3$ are the smallest and largest eigenvalues,
respectively. Note that this definition differs from that
of~\citet{Bos2012}.

\begin{figure*} 
  \centering 
  {\includegraphics[type=png,ext=.png,read=.png,width=0.48\textwidth]{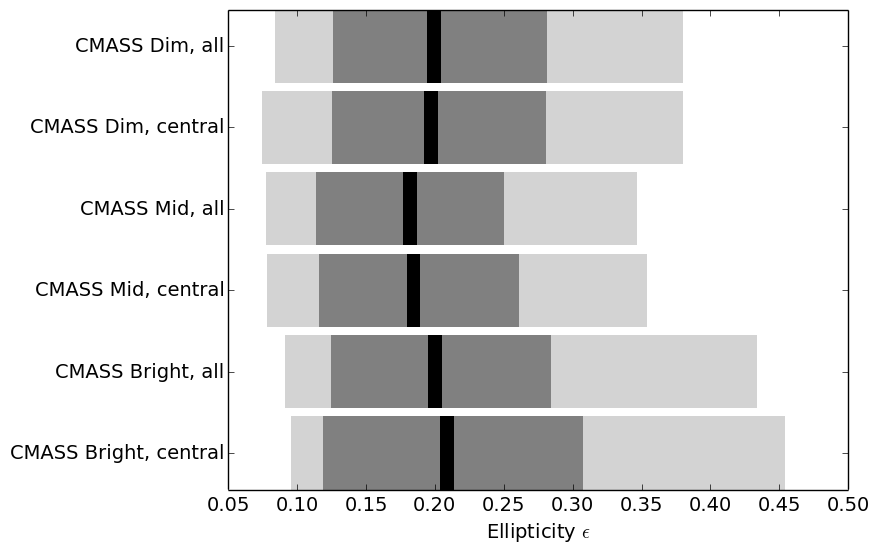}}
  {\includegraphics[type=png,ext=.png,read=.png,width=0.48\textwidth]{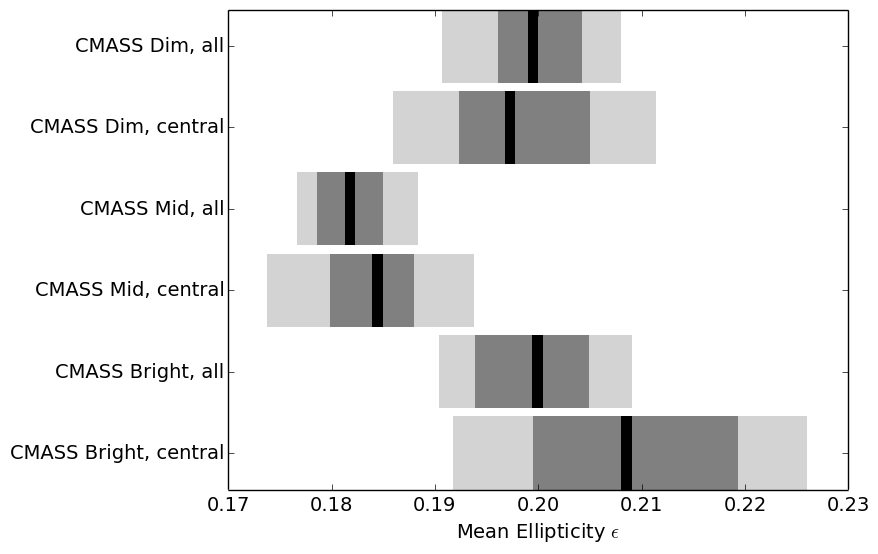}}
  \caption{
           Ellipticity distributions $\epsilon$ (Eq.~\ref{eq:ellip}) with
           68\% (dark grey) and 95\% (light grey) ranges 
           for each sample (left),
           and mean ellipticities with 1$\sigma$ and 2$\sigma$ uncertainties
           on the mean (right). For the mean ellipticities (right), error
           bars are calculated with a bootstrap method.
          }
\label{fig:ellip_data}
\end{figure*}

Figure~\ref{fig:ellip_data} shows the distribution of ellipticities
for each of our samples. 
The ellipticities for all the samples are remarkably consistent, 
with means $\sim 0.2$ and a slight skew in the distribution favoring 
slightly more elliptical voids. The \emph{central} catalog 
of the \emph{CMASS Bright} voids contains a few highly elliptical
voids. While similar voids exist in all samples, the limited 
number of total voids in this sample leads to a highly 
lopsided distribution.
In the same figure, we also
show the mean ellipticity and the standard
error on the mean (i.e., $\sigma/{N_v}$, where $\sigma$ is the standard
deviation and $N_v$ is the number of voids) for each sample.
The means broadly agree, with the \emph{CMASS Mid} sample favoring 
slightly more spherical voids. 
The ellipticities in the \emph{central} catalogs are 
different than those in the \emph{all} catalogs.
We will see below that the mean 
ellipticity is sensitive to the distribution of void sizes in a 
particular sample. Since the mask affects the void size distribution 
in a non-trivial way for each sample, depending on the relative 
surface-to-volume ratio in the sample, we should not be surprised 
to measure slightly different mean void ellipticities.
Fortunately, this does not appear to be a large effect 
from sample to sample, and we conclude that we only
need to understand mask effects 
for a particular survey geometry, not for individual 
volume-limited samples within that survey. 

\begin{figure*} 
  \centering 
  {\includegraphics[type=png,ext=.png,read=.png,width=0.32\textwidth]{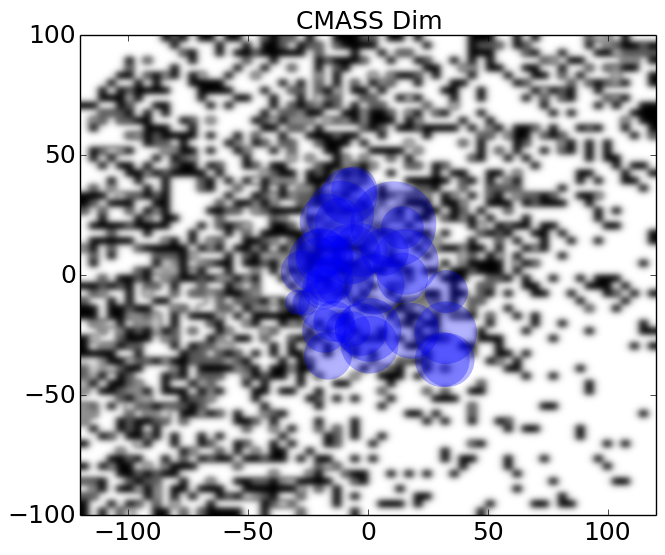}}
  {\includegraphics[type=png,ext=.png,read=.png,width=0.32\textwidth]{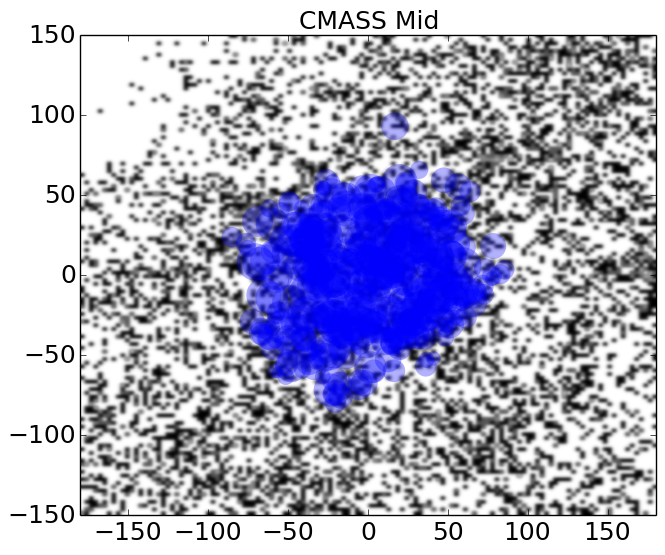}}
  {\includegraphics[type=png,ext=.png,read=.png,width=0.32\textwidth]{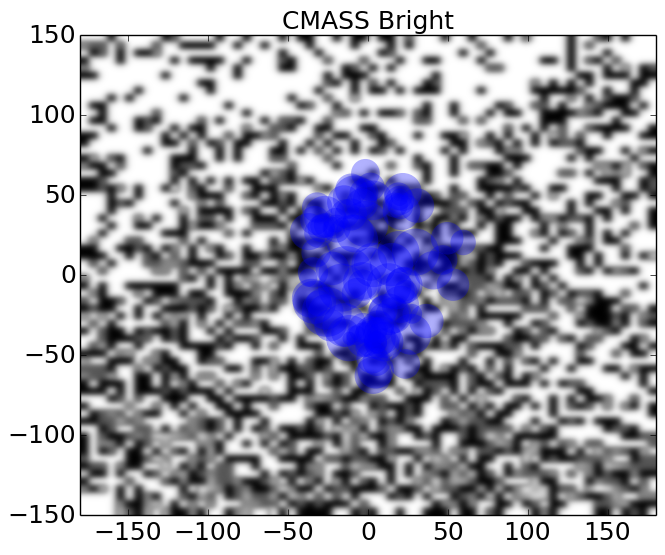}}
  \caption{
           Void and
           galaxy density slices.
           We select three locations in each 
           sample centered on a representative void.
           We show voids from \emph{CMASS Dim} (left),
           \emph{CMASS Mid} (middle), and \emph{CMASS Bright} (right).
           We represent the void member galaxies as small circles with 
           radii equal to the effective radii of their corresponding Voronoi 
           cells.
           The width of each galaxy slice along the $z$-axis is
           the entire sample for \emph{CMASS Dim} and 300 \hmpc~for 
           \emph{CMASS Mid} and \emph{CMASS Bright}.
           To avoid unnecessary overlap from projection 
           we take void particles from a thinner slice: 
           10 \hmpc~for \emph{CMASS Dim}, 75 \hmpc~for \emph{CMASS Mid}, 
           and 150~\hmpc~for \emph{CMASS Bright}.
           The galaxies are binned into pixels with the number of pixels 
           varied to best highlight the surrounding structure:
           64 bins for \emph{CMASS Dim}, 128 for \emph{CMASS Mid}, and
           64 for \emph{CMASS Bright}.
           The projected galaxy density is colored from 
           $0.0$ (white) to $1.5$ (black) and is shown on a logarithmic 
           scale.
           Axes are marked in units of~\hmpc.
          }
\label{fig:slice}
\end{figure*}

In Figure~\ref{fig:slice} we give a visual impression of some 
of the identified voids. 
We chose the particular slices randomly
but selected a representative sample from
the range of scales in the void catalog.
We represent the selected void as collections of overlapping circles, 
where each circle is a void member galaxy with radius equal to the 
effective radius of each Voronoi cell.
We overplot these circles on slices from the galaxy 
distribution. 
We see that voids at all scales indeed sit within underdensities in the 
galaxy distribution, though in some samples the sparsity makes it 
difficult to clearly distinguish the surrounding walls and filaments.
However, our analysis in~\citet{Sutter2013a} and~\citet{Hamaus2014} 
show that these objects share common features with voids in high-density 
surveys.

\begin{figure*} 
  \centering 
  {\includegraphics[type=png,ext=.png,read=.png,width=0.48\textwidth]{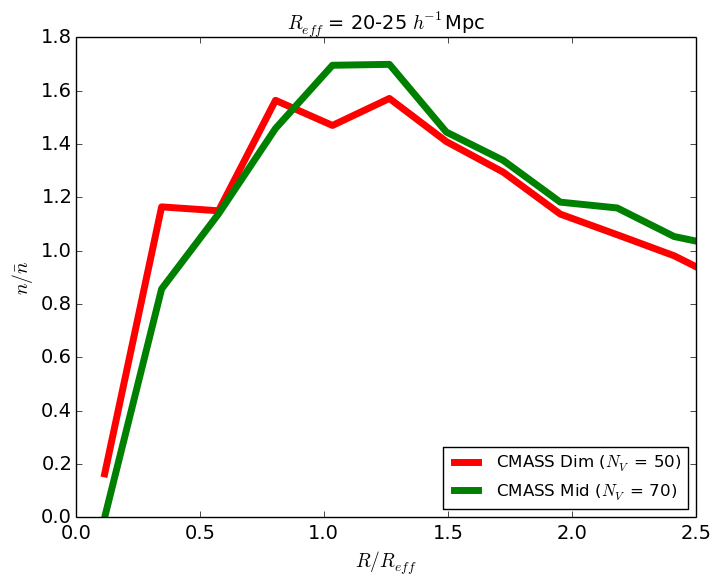}}
  {\includegraphics[type=png,ext=.png,read=.png,width=0.48\textwidth]{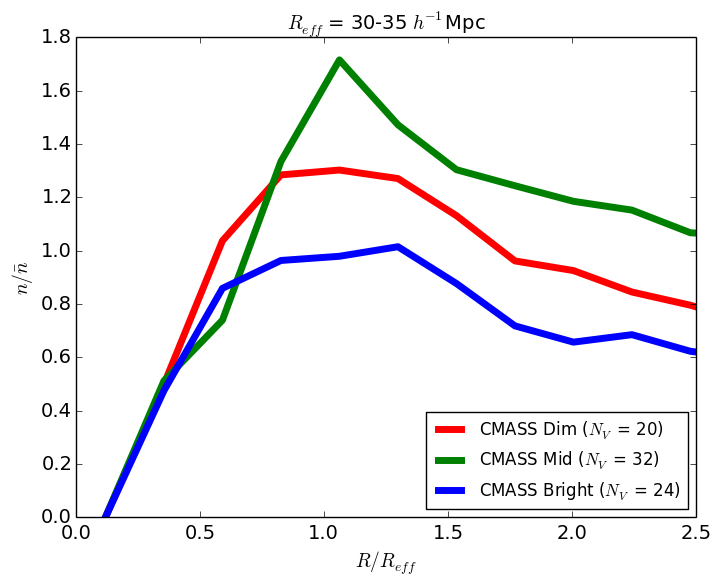}}
  {\includegraphics[type=png,ext=.png,read=.png,width=0.48\textwidth]{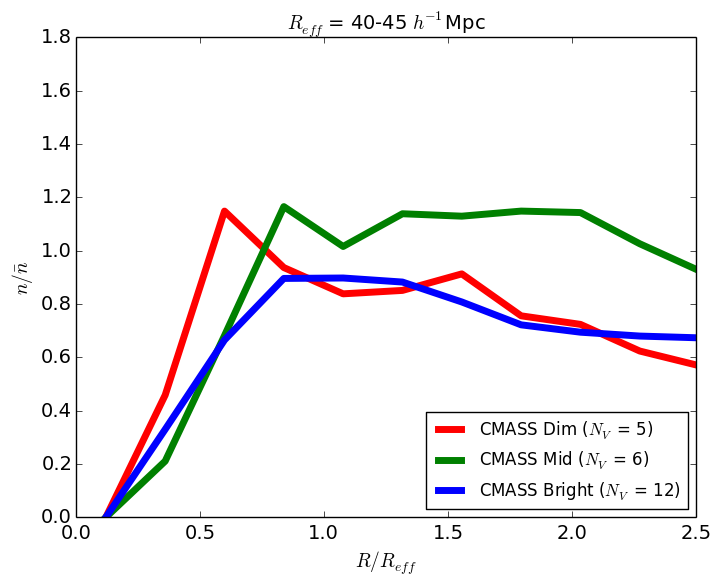}}
  {\includegraphics[type=png,ext=.png,read=.png,width=0.48\textwidth]{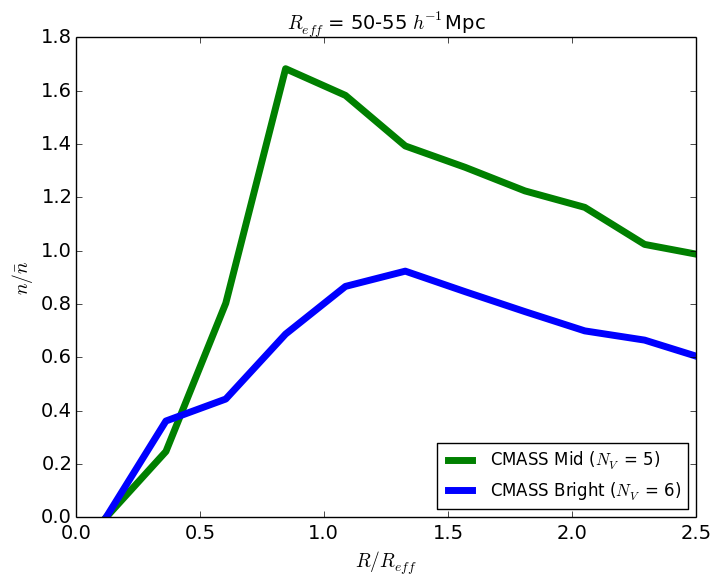}}
  \caption{
           One-dimensional radial density profiles of stacked voids.
           Each profile is normalized to the mean number density $\bar{n}$
           of that sample and $R_{\rm eff}$ corresponds to the median
           void size in the stack.
           We only show profiles from stacked \emph{central} voids.
           Void profiles do not necessarily reach the mean 
           density because of the influence of boundary particles 
           and empty regions outside the survey volume.
           The caption lists the number of voids stacked in each profile.
          }
\label{fig:1dprofile_data}
\end{figure*}

Figure~\ref{fig:1dprofile_data} shows one-dimensional 
radial density profiles of stacked voids in each sample. 
To compute the profiles, we take all voids in a
sample of a
given size range (e.g., $20-25$ \hmpc), align all their macrocenters,
and measure the density in thin spherical shells. We normalize each
density profile to the mean number density of the sample and show
all profiles as a function of relative radius, $R/R_{\rm eff}$.
We only show profiles from the \emph{central}
voids of the samples. 
We show four stacks: 20-25, 30-35, 40-45, and 50-55 \hmpc.

The smallest stack, 20-25 \hmpc, is very close to the mean galaxy 
separation for the samples, and this manifests in an extremely steep
profile. 
While these small voids may be unreliable due to Poisson shot 
noise, we do not assess statistical significance in this work. 
We believe
that the common technique used with watershed algorithms (comparing to voids in
a Poisson distribution of equal number density) is inadequate because small
voids tend to appear in higher density regions, and thus are more likely 
to be real voids than one would expect. We are currently developing 
a more robustsignificance criteria based on Bayesian analysis of constrained realizations of
a given sample.

As we progress to larger stacks, the profiles become more 
shallow and the overdense region surrounding the voids becomes 
less pronounced, as also seen in~\citet{Ceccarelli2013}.
However, as before in the voids of SDSS DR7~\citep{Sutter2012a}, 
we see a qualitatively universal profile across all void 
sizes and redshift ranges: an underdense center, a steep power-law 
slope at the wall of the void, a slightly overdense ``compensation'', 
and a steady declining to the mean density. 
We see here that there is a larger difference among the 
samples than in the DR7 voids. A significant reason for this is the 
smaller survey area: while we guarantee \emph{central} voids to 
sit completely within the survey, these profiles extend 
beyond the void effective radius. Since the survey area is 
relative small, the profile quickly reaches into volumes 
beyond the survey mask. There are also boundary particles at the edges.
Depending on the relative density of the boundary particles and the 
location of stacked voids relative to the boundary we see 
different profiles at larger radii.
Within $R_{\rm eff}$, we see strong consistency among the samples, 
as expected.
The number of voids in the stack strongly affects the smoothness of 
the profile.
In particular, the 50-55 \hmpc~stack in \emph{CMASS Mid} contains 
only 5 voids, which leads to a highly irregular profile shape.

\section{Voids in Mocks}
\label{sec:voidmocks}

\subsection{Mock Galaxy Populations}

We take several avenues for comparison to the voids in the survey 
data. For the first set of mocks, we compute a single
\lcdm~dark matter $N$-body simulation, extract halos from the simulation, 
and use the halos positions and masses as inputs for an HOD model.
For the simulation we use the {\tt 2HOT} code, 
an adaptive treecode N-body method whose operation count
scales as $N \log N$ in the number of particles~\citep{warren13}.
Accuracy and error behavior have been improved significantly for
cosmological volumes through the use of a technique to subtract the
uniform background density, as well as using a compensating smoothing
kernel for small-scale force softening~\citep{dehnen01}.  We use a
standard symplectic integrator~\citep{quinn97} and an efficient
implementation of periodic boundary conditions using a high-order
($p=8$) multipole local expansion. We adjust the error tolerance
parameter to limit absolute errors to 0.1\% of the rms peculiar
acceleration. As an example, a
complete $4096^3$ particle simulation requires about
120 wall-clock hours using 12,000 CPU cores. Initial conditions were
generated using a power spectrum calculated with {\tt CLASS}~\citep{blas11}
and realized with a modified version of {\tt 2LPTIC}~\citep{crocce06}.

This particular simulation assumed Planck first-year cosmological
parameters~\citep{Planck2013}. The box size was 4\hgpc~on a side
and contained $4096^3$ particles, giving a particle mass resolution
of $7.36 \times 10^{10} \hmsol$.
All analysis in this work used a single snapshot at
$z=0.53$.
We identified halos in the simulation volume using the 
 {\tt Rockstar} halo finder~\citep{behroozi13}, a
six-dimensional phase-space plus
time halo finder, to identify spherical overdensity (SO) halos at 200
times the background density.  We use the default Rockstar parameters,
except for requiring strict SO masses which includes unbound particles
and particles which may exist outside of the FOF group for the halo.

We produce galaxy catalogs from the halo population
using the code described in~\citet{Tinker2006} and the HOD model
described in~\citet{Zheng2007}.
HOD modeling assigns central and satellite galaxies to a dark matter
halo of mass $M$ according to a parametrized distribution.
In the case of the~\citet{Zheng2007} parametrization, the
mean number of central galaxies is given by
\begin{equation}
\left\langle N_{\rm cen}(M)\right\rangle = \frac{1}{2} \left[
1 + \erf \left(\frac{\log M - \log M_{\rm min}}{\sigma_{\log M}}\right)
\right]
\end{equation}
and the mean number of satellites is given by
\begin{equation}
\left< N_{\rm sat}(M)\right> = \left\langle N_{\rm cen}(M) \right\rangle
\left( \frac{M-M_0}{M_1'}\right)^\alpha,
\end{equation}
where $M_{\rm min}$, $\sigma_{\log M}$, $M_0$, $M_1'$, and $\alpha$
are free parameters that must be fitted to a given survey.
The probability distribution of central galaxies $P(N_{\rm cen} | \left< 
N_{\rm cen} \right>)$ is a nearest-integer
distribution, and satellites follow a Poisson
$P(N_{\rm sat} | \left< N_{\rm sat} \right>)$.
Central galaxies are given the peculiar velocities of the host halo,
and satellite galaxies are given an additional random 
velocity 
drawn from a Maxwellian distribution with the halo velocity 
dispersion.
Using the HOD parameters found in~\citet{Manera2013}
($\sigma_{\log M} = 0.596$, $M_0 = 1.2\times10^{13} \hmsol$, $M_1' = 10^{14} \hmsol$, $\alpha = 1.0127$, and $M_{\rm min}$ chosen to fit the mean number density of our sample)
 we generate
the mock galaxy population. Although the HOD fitting of~\citet{Manera2013}
assumed a slightly different cosmology, we found that this did not 
affect our results. To simplify comparison, we will only target the 
\emph{CMASS Mid} sample.

We note that we only fit the mean density of the galaxy sample; we 
make no attempt to model the variation of the number density as a function 
of redshift. However, even with this restriction, we will see below that 
we find excellent agreement between the mocks and data for all void 
statistics. While ignoring the density variation 
 may not be adequate for precise 
modeling, at the current level of statistical uncertainty we only wish 
to make an initial comparison, and save a more detailed 
treatment for future work.

For the full-volume simulation analysis, we use the entire 
three-dimensional volume of the simulation box and
perturb each galaxy according to 
its peculiar velocity. 
We call this full-volume set of mock 
galaxies the \emph{N-body Mock} sample.
Even though we analyze only 
a single realization, the large volume produces over $90,000$ 
voids.

To provide a more direct comparison to the data and to understand 
the effects of the mask on void properties, we apply the 
same survey geometry to mock galaxies as is used in the SDSS DR9 
samples. Instead of placing galaxies in redshift space along 
the $z$-axis of the simulation box as we do above, we place an 
observer at the center of the volume and measure each galaxy position
as its radial distance from that observer. We also perturb the 
galaxies according to the peculiar velocities in all directions.
We project all galaxies onto the sky and apply the mask in
Figure~\ref{fig:mask}. Since we wish to compare solely to the \emph{CMASS Mid}
sample, we only take mock galaxies within $0.43< z < 0.5$. 
For void finding we treat this sample exactly as data: we deploy 
boundary particles and we build \emph{all} and \emph{central} 
catalogs of voids. While it is very expensive to compute multiple 
realizations with this volume and resolution, we can take advantage 
of the relative narrowness of the current DR9 survey by 
rotating our mask orientation within the same simulation volume.
We take 5 separate orientations.
Even though these samples are not quite independent, and we are likely to 
not capture enough of the very largest voids due to finite-volume
effects, this technique still provides a good proxy for 
multiple realizations and allows us to gauge the range 
of void statistics predicted from simulations.
We call these samples \emph{Masked N-body Mock}.
Taken together, we find $\sim 3,300$ voids in the \emph{all} samples 
of all orientations, and $\sim 1300$ voids in the \emph{central} 
samples.

Since high-resolution $N$-body simulations are very expensive, and 
the galaxy correlation analysis of~\citet{Reid2012} required many 
mocks for estimating shot noise, the BOSS team produced many 
mock catalogs based on second-order Lagrangian perturbation theory (2LPT)
with a WMAP 7-year cosmology
and the same HOD prescription described above~\citep{Manera2013}.
We used 98 of these publicly-available mock catalogs to validate 
our $N$-body simulation results and to further estimate the 
uncertainties in the predicted void statistics. 
We denote these mock samples as \emph{Masked 2LPT Mocks} and process them 
identically to real data.
With the 98 mocks we find over $56,000$ voids in the \emph{all} sample 
and over $25,000$ in the \emph{central} sample.
However, these mocks 
were only made available with the survey mask already in place; 
thus we will only use our \emph{N-body Mocks} for interpreting the 
effects of the mask.

Finally, to evaluate the significance of our ellipticity measurements,
we identifies voids in a single random realization provided by the CMASS 
team. In this realization, galaxies are randomly distributed with 
Poissonian statistics within the survey volume. 

Table~\ref{tab:voidsamples_mock} summarizes the mock samples used in this work,
their respective volumes, the number of independent mock samples,
and the total number of voids identified 
in each sample.

\begin{table}
\centering
\caption{Summary of voids in mocks.}
\tabcolsep=0.11cm
\footnotesize
\begin{tabular}{ccccc}
  Sample Name & Volume & \# Mocks & $N_{\rm voids}$ & \\
   & (\hgpccubed) &  & \\
  \hline  \hline
N-body Mock & 64.00 & 1 & 91711\\
Masked N-body Mock, all & 0.53 & 5 & 3336\\
Masked N-body Mock, central & 0.53 & 5 & 1313\\
Masked 2LPT Mock, all & 0.53 &  98 & 56096\\
Masked 2LPT Mock, central & 0.53 & 98 & 25299\\
\hline
\end{tabular}
\label{tab:voidsamples_mock}
\end{table}

\subsection{Comparison to Data}

\begin{figure} 
  \centering 
  {\includegraphics[type=png,ext=.png,read=.png,width=\columnwidth]{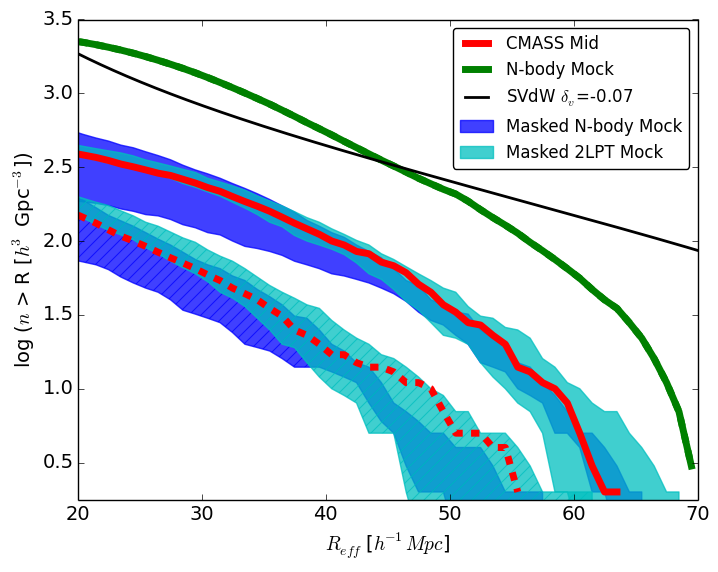}}
  \caption{
          Comparison of cumulative void number function between mocks 
          and data.
          The solid (dotted) line corresponds to the \emph{all} 
          (\emph{central}) void population of the \emph{CMASS Mid}
          galaxy sample. The green solid line shows the number function of 
          the full-volume \emph{N-body Mock} sample. 
          The light blue shaded region shows the full 
           range of number functions from
          the 98 \emph{Masked 2LPT Mock} runs, while the dark blue
           shaded region shows 
          the full range of the 5 \emph{Masked N-body Mock} runs. For the 
          shaded regions, solid indicates \emph{all} voids and 
          hatched indicates \emph{central} voids.
          The solid black line is the theoretical expectation from the 
          ~\citet{Sheth2004} number function with 
          $\delta_v=-0.07$.
          }
\label{fig:numberfunc_mockvsdata}
\end{figure}

Our first point of comparison is the cumulative number function.
In Figure~\ref{fig:numberfunc_mockvsdata} we compare the number function 
of voids in the \emph{CMASS Mid} data sample to all our mocks.
First, the unmasked \emph{N-body Mock} simulation hosts roughly 
three times as many voids per unit volume than the data, even 
though they have similar galaxy populations. This occurs at all scales, 
though there are $\sim 4$ times as many small voids in the unmasked 
mock as in the data.
We can understand this disruption by considering the effects 
of the mask on a particular void: it will slice the void, making it 
appear as a smaller voids. So after the mask is applied large 
voids become medium voids, medium voids become small voids, and so on.
Since the number function falls steeply with void size, 
if we consider a given void size range, then there are far fewer 
voids being added to that range (by being sliced and becoming smaller) 
than voids that are lost to smaller ranges. This leads to a systematic 
reduction in the number density of voids.
For larger survey areas and more complete coverage, we expect 
less drastic impacts.

In the same figure we plot the theoretical number function 
of~\citet{Sheth2004}, which was derived from an excursion set 
formalism. As found in~\citet{Sutter2013a}, the best match to 
voids in low-density galaxy surveys comes from adjusting the 
``void parameter'' $\delta_v$ to $-0.07$. While the 
number function roughly agrees with the order of magnitude of the 
full \emph{N-body Mock} void population, it does 
not fall off as steeply as in the mocks, though this might be 
influenced by finite-volume effects. Still, the correspondence 
of these curves shows that theoretical modeling can qualitatively match 
unmasked void populations, but further adjustments must be 
made to match void statistics from masked volumes.

When we apply the DR9 survey geometry to our mock catalogs, we see 
excellent agreement in the number functions for both \emph{all} 
and \emph{central} catalogs. The \emph{2LPT} mocks 
show much less variance at small void sizes and reach larger void sizes
than the \emph{N-body} mocks, since there are many more 
\emph{2LPT} simulations and they are each drawn from independent 
realizations. The different orientations of the \emph{N-body} mocks 
are limited by the cosmic variance of a single simulation, 
so it is not surprising that we are not able to match the 
void number function at the largest void sizes. 
However, with the \emph{2LPT} mocks, we are able to 
capture the very largest voids.~\citet{Tavasoli2013} first pointed 
out a potential tension between \lcdm~model and data, 
but we see no indication of this tension in our results.

\begin{figure*} 
  \centering 
  {\includegraphics[type=png,ext=.png,read=.png,width=0.48\textwidth]{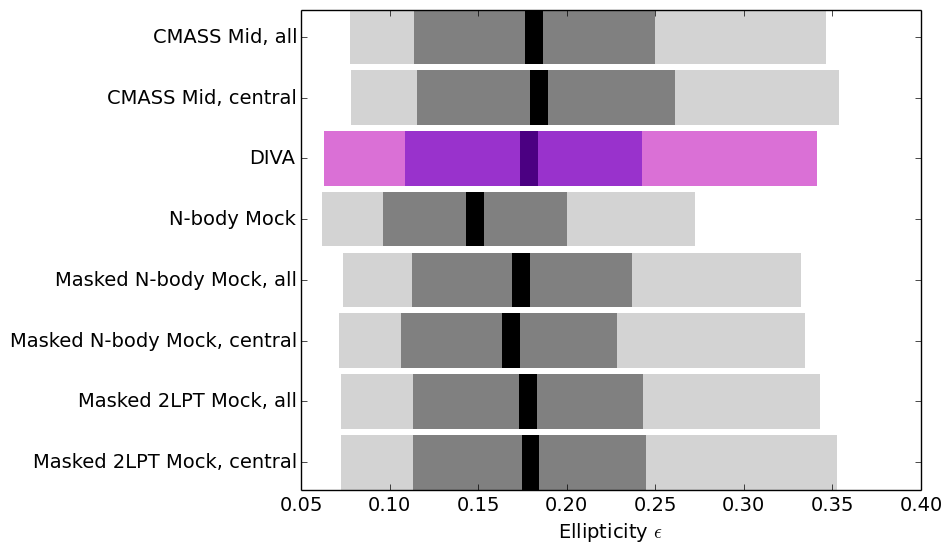}}
  {\includegraphics[type=png,ext=.png,read=.png,width=0.48\textwidth]{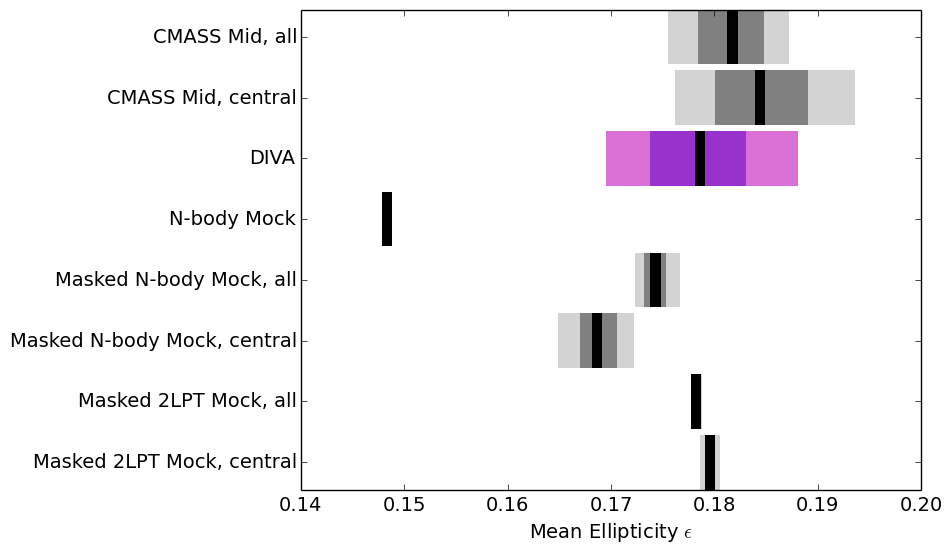}}
  \caption{
           Comparing the ellipticity distributions (left) and mean 
           ellipticities with their uncertainties 
           (right) of the \emph{CMASS Mid} sample to the 
           various mocks. Error bars are the same as indicated for 
           Figure~\ref{fig:ellip_data}. 
           The violet-colored distribution is the theoretical 
           expectation from {\tt DIVA}~\citep{Lavaux2010} with 
           the rescaling parameter $\alpha$ set to $0.25$.
           {\tt DIVA} predicts the ellipticity distribution from
           the size distribution of voids in the \emph{CMASS Mid} 
           \emph{central}
           sample.
           }
\label{fig:ellip_comparemid}
\end{figure*}

Figure~\ref{fig:ellip_comparemid} shows the ellipticity distributions 
and mean ellipticities, identically to Figure~\ref{fig:ellip_data}. 
In this plot, we compare \emph{CMASS Mid} ellipticities to those 
found in mock samples.
Voids in all mock and data samples 
tend to be more elliptical than voids drawn from Poissonian distributions 
(which give $\epsilon \sim 0.12$)
, confirming the analysis of~\citet{Sutter2013b}, which showed that
ellipticity is a robust measurement even in low-density samples such 
as SDSS DR9.

As we saw earlier for the cumulative number functions, we find 
agreement between data and theoretical expectations in the form of mock 
catalogs \emph{only} when we apply the mask: then the distributions are 
nearly identical. The \emph{all} void samples agree among all the 
masked mocks and data. The mean ellipticities of the \emph{central} 
\emph{Masked N-body Mock} 
are slightly more than 2$\sigma$ discrepant than the data, 
while the \emph{Masked 2LPT Mocks} have much better agreement with the 
data.
The differences between the \emph{N-body} and \emph{2LPT} 
mock populations are not surprising since they have slightly 
different cosmologies, and the ellipticity is a very sensitive probe 
of cosmological parameters~\citep{Biswas2010,Bos2012}.
Also, the \emph{2LPT} Mocks cover multiple realizations, whereas the 
\emph{N-body} mocks are restricted to a single simulation.
With the exception 
of the \emph{Masked N-body Mock}, we see the same relationship between 
\emph{all} and \emph{central} ellipticities as in the data: 
voids in \emph{central} samples less spherical.
Indeed, the $90,000$ voids in all the \emph{2LPT} mocks reduce 
the uncertainties to such a degree that this difference 
is easily distinguishable.

\begin{figure} 
  \centering 
  {\includegraphics[type=png,ext=.png,read=.png,width=\columnwidth]{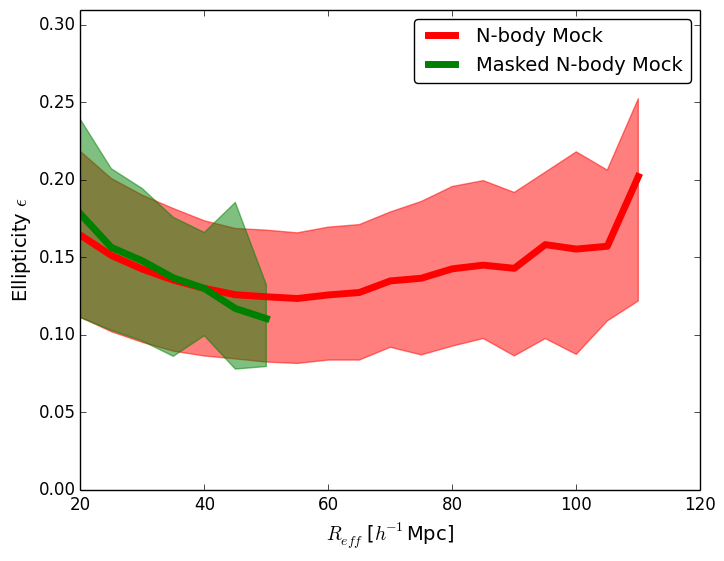}}
  \caption{
          Distribution of ellipticities $\epsilon$ 
          in small bins of effective radius $R_{\rm eff}$
          in the masked (green) and unmasked (red) \emph{N-body Mock}
          simulations. The solid line shows the mean, and the 
          shaded region is the $1\sigma$ interval.
          }
\label{fig:ellipradius}
\end{figure}

Voids in the data and 
masked mocks are more elliptical than in the full-volume simulation.
We can understand the impact of the mask by examining the relationship 
between ellipticity and void size, as we show for the masked and 
unmasked \emph{N-body} mocks in Figure~\ref{fig:ellipradius}.
The cause of the shift in ellipticities when masking data is 
readily apparent: larger voids tend to be more spherical, 
and their exclusion from the masked catalogs increases 
the overall mean ellipticity.
The uncertainties in the \emph{N-body} mocks are too large to 
distinguish any differences.

We derive the theoretical ellipticity distribution 
from {\tt DIVA}~\citep{Lavaux2010} with the rescaling 
parameter $\alpha=0.25$, as discussed in the analysis 
of~\citet{Sutter2013a}. {\tt DIVA} requires a void size distribution 
as input, and we take the actual distribution from the \emph{CMASS Mid}
\emph{central} sample (i.e., Figure~\ref{fig:numberfunc_data}).
This choice of rescaling parameter $\alpha$ provided good agreements 
with mock void populations, and again here we see agreement 
with data, indicating a relationship between the sizes of voids 
identified with the watershed transform and their dynamical cores. 

\begin{figure*} 
  \centering 
  {\includegraphics[type=png,ext=.png,read=.png,width=0.48\textwidth]{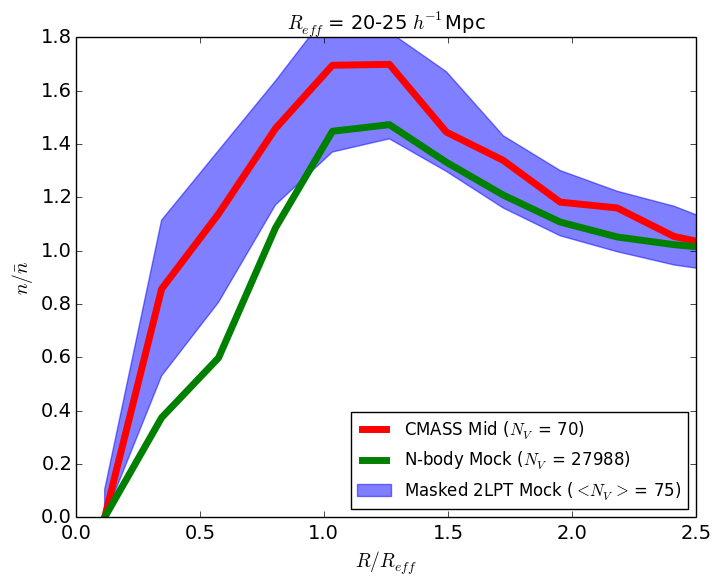}}
  {\includegraphics[type=png,ext=.png,read=.png,width=0.48\textwidth]{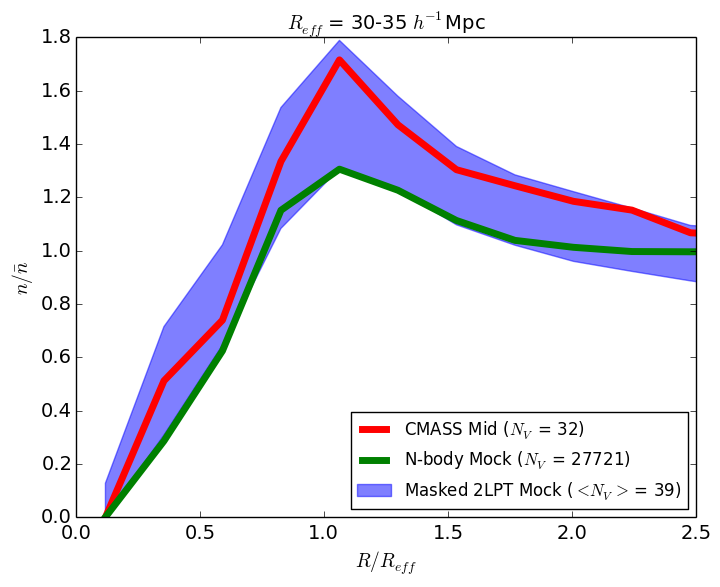}}
  {\includegraphics[type=png,ext=.png,read=.png,width=0.48\textwidth]{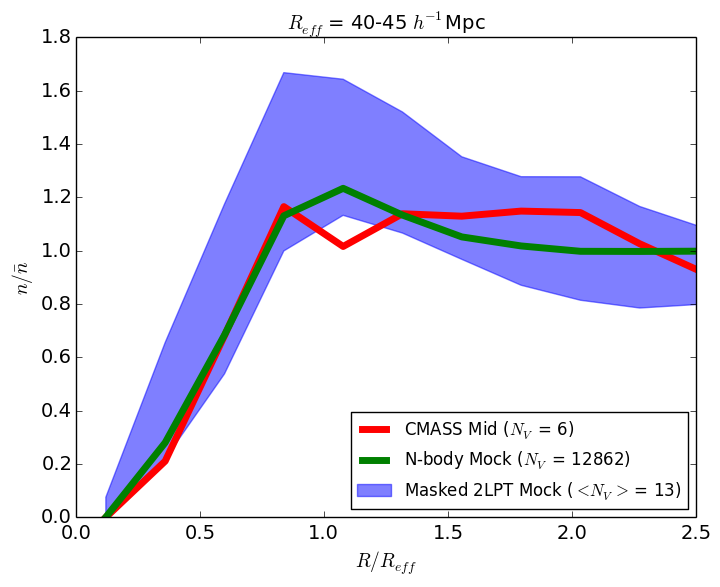}}
  {\includegraphics[type=png,ext=.png,read=.png,width=0.48\textwidth]{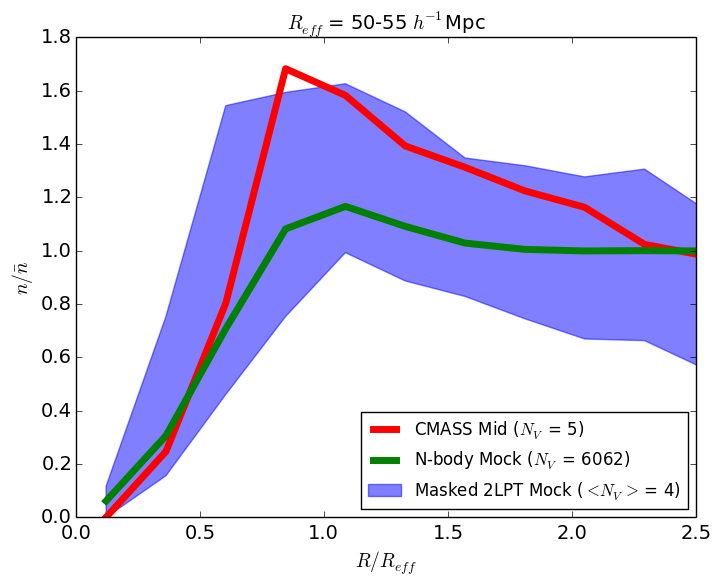}}
  \caption{
          Comparison of one-dimensional radial density profiles. The void
          stacks and normalizations are the same as in
          Figure~\ref{fig:1dprofile_data}. The solid red lines are profiles
          from \emph{CMASS Mid}, the solid green lines are from 
          the full-volume unmasked \emph{N-body Mock}, while the shaded 
          blue region shows the full range of profiles from 
          the 98 realizations
          of the \emph{Masked 2LPT Mock} simulations.
          The caption lists the number of voids stacked in each profile, 
          except for \emph{Masked 2LPT Mock}, where the caption shows the 
          mean number of voids in each realization for that stack.
          The number of excursions of the data from the mocks is consistent
          with expectations.
          }
\label{fig:1dprofile_comparemask}
\end{figure*}

Interestingly, we see almost no significant differences between 
masked and unmasked radial density profiles, as we show in
Figure~\ref{fig:1dprofile_comparemask}. Here we compare the 
unmasked \emph{N-body Mock} and the \emph{Masked 2LPT Mocks} 
to the \emph{CMASS Mid} data sample.
We do not plot stacks from the \emph{Masked N-body Mock} 
since they simply overlap the exisiting curves but have larger scatter.
The stacks shown here are the same as before in 
Figure~\ref{fig:1dprofile_data}. With very few exceptions, 
profiles from both the data and the full-volume mock
sit within the range of profiles constructed in the different 
realizations of the masked mock. This agreement persists at 
all void scales. 
The two visible excursions from the mocks is consistent with 
statistical expectations.
Since radial density profiles by definition probe the interiors 
and immediate surroundings of voids, they naturally are more 
robust against distortions due to the mask. Also, the process of 
stacking turns sets of elliptical voids with different orientations 
into a roughly spherical shape, regardless of the distribution 
of the individual ellipticities~\citep{LavauxGuilhem2011,Pisani2013}.
Thus any changes to the mean ellipticity due to the mask will not 
change the radial profiles significantly.

With the enormous number of voids in the \emph{N-body Mock} sample, we 
begin to see 
the transition from over- to under-compensated voids~\citep{Hamaus2013}.
 However, this 
transition is not clear in the data due to the lack of voids at these 
extreme scales.

\section{Conclusions}
\label{sec:conclusions}

We have constructed a void catalog from the SDSS DR9 
CMASS spectroscopic galaxy survey. Combined with the voids 
from DR7, this is the largest void catalog to date.
We have used previously-established methods for removing voids 
near survey boundaries to produce a catalog of voids with a 
fair distribution of shapes.
This catalog also provides the most distant voids ever detected, 
extending our knowledge of galaxy underdensities to 
redshift $z=0.7$. Our voids have effective radii between 20 and 70 \hmpc,
and exhibit the same qualitatively universal radial density profile 
seen in earlier works~\citep{Sutter2012a,Ceccarelli2013}. 
We used Halo Occupation Distribution modeling to produce mock 
catalogs for comparison purposes.

We find that the effects of the mask are highly non-trivial and 
can depend strongly on the relative surface area of the mask 
compared to the volume, 
the depth of the survey, the number of internal holes, and the 
detailed shape of the boundary. 
Cosmological statistics based on global void properties, 
such as number functions and ellipticity distributions, 
are especially vulnerable to the properties of the mask. 
However, we find that probes based on void \emph{interiors}, such as radial 
density or velocity profiles, 
are generally more robust. Also, cosmological applications which 
depend on the statistical isotropy of voids, such as the 
Alcock-Paczynski test~\citep{Sutter2012b}, are 
resilient. In cases such as this, while the mask may change 
the average void shape or size, as long as the mask preserves a 
uniform sampling of their orientations then the methods are sound.

\citet{Furlanetto2006} and~\citet{Tinker2009} hypothesized that the
so-called ``void phenomenon'' of Peebles (2001) can be explained by
galaxy bias. This has led to a general discussion in the literature about
potential discrepancies between the voids in theory and voids in data. For the void definition we adopt,
abundances, ellipticity distributions, and radial profiles all indicate that
voids in simulations have the same sizes, shapes, and interior contents as
observed voids once galaxy bias, sparsity, and survey masks are accounted for.
Overall, we find no significant discrepancies between observations of voids in
the SDSS DR9 and \lcdm~mocks.

We have made all the SDSS DR9 voids, as well as all mock 
catalogs used in this work, publicly available online
at the Public Cosmic Void Catalog
at \footnotesize{http://www.cosmicvoids.net}.

\section*{Acknowledgments}

PMS and BDW acknowledge
support from NSF Grant NSF AST 09-08693 ARRA. BDW 
acknowledges funding from an ANR Chaire d'Excellence (ANR-10-CEXC-004-01),
the UPMC Chaire Internationale in Theoretical Cosmology, and NSF grants AST-0908
902 and AST-0708849.
GL acknowledges support from CITA National Fellowship and financial
support from the Government of Canada Post-Doctoral Research Fellowship.
Research at Perimeter Institute is supported by the Government of Canada
through Industry Canada
 and by the Province of Ontario through the Ministry of Research and
Innovation.
DW acknowledges support from NSF Grant AST-1009505.
This work made in the ILP LABEX (under reference ANR-10-LABX-63) was supported by French state funds managed by the ANR within the Investissements d'Avenir programme under reference ANR-11-IDEX-0004-02. 
We would like to thank Nico Hamaus for useful discussions.

\footnotesize{
  \bibliographystyle{mn2e}
  \bibliography{catalogdr9}

\begin{thebibliography}{}

\bibitem[\protect\citeauthoryear{{Ahn} et~al.,}{{Ahn}  et~al.}{2012}]{Ahn2012}
{Ahn} C.~P.,  et~al., 2012, \apjs, 203, 21

\bibitem[\protect\citeauthoryear{Alcock \& Paczynski}{Alcock \&
  Paczynski}{1979}]{Alcock1979}
Alcock C.,  Paczynski B.,  1979, Nature, 281, 358

\bibitem[\protect\citeauthoryear{{Bassett} \& {Hlozek}}{{Bassett} \&
  {Hlozek}}{2010}]{Bassett2010}
{Bassett} B.,  {Hlozek} R.,  2010, {Baryon acoustic oscillations}.
p.~246

\bibitem[\protect\citeauthoryear{{Baugh}, {Benson}, {Cole}, {Frenk} \&
  {Lacey}}{{Baugh} et~al.}{2003}]{Baugh2003}
{Baugh} C.~M.,  {Benson} A.~J.,  {Cole} S.,  {Frenk} C.~S.,    {Lacey} C.,
  2003, in {Bender} R.,  {Renzini} A.,  eds, The Mass of Galaxies at Low and
  High Redshift {The Evolution of Galaxy Mass in Hierarchical Models}.
p.~91

\bibitem[\protect\citeauthoryear{{Beck}, {Hanasz}, {Lesch}, {Remus} \&
  {Stasyszyn}}{{Beck} et~al.}{2013}]{Beck2013}
{Beck} A.~M.,  {Hanasz} M.,  {Lesch} H.,  {Remus} R.-S.,    {Stasyszyn} F.~A.,
  2013, \mnras, 429, L60

\bibitem[\protect\citeauthoryear{Behroozi, Wechsler \& Wu}{Behroozi
  et~al.}{2013}]{behroozi13}
Behroozi P.~S.,  Wechsler R.~H.,    Wu H.,  2013, The Astrophysical Journal,
  762, 109

\bibitem[\protect\citeauthoryear{Benson, Hoyle, Torres \& Vogeley}{Benson
  et~al.}{2003}]{Benson2003}
Benson A.~J.,  Hoyle F.,  Torres F.,    Vogeley M.~S.,  2003, \mnras, 340, 160

\bibitem[\protect\citeauthoryear{{Berlind} \& {Weinberg}}{{Berlind} \&
  {Weinberg}}{2002}]{Berlind2002}
{Berlind} A.~A.,  {Weinberg} D.~H.,  2002, \apj, 575, 587

\bibitem[\protect\citeauthoryear{Biswas, Alizadeh \& Wandelt}{Biswas
  et~al.}{2010}]{Biswas2010}
Biswas R.,  Alizadeh E.,    Wandelt B.,  2010, \prd, 82

\bibitem[\protect\citeauthoryear{Blas, Lesgourgues \& Tram}{Blas
  et~al.}{2011}]{blas11}
Blas D.,  Lesgourgues J.,    Tram T.,  2011, Journal of Cosmology and
  Astroparticle Physics, 2011, 034

\bibitem[\protect\citeauthoryear{{Bos}, {van de Weygaert}, {Dolag} \&
  {Pettorino}}{{Bos} et~al.}{2012}]{Bos2012}
{Bos} E.~G.~P.,  {van de Weygaert} R.,  {Dolag} K.,    {Pettorino} V.,  2012,
  ArXiv e-prints: 1205.4238

\bibitem[\protect\citeauthoryear{{Ceccarelli}, {Herrera-Camus}, {Lambas},
  {Galaz} \& {Padilla}}{{Ceccarelli} et~al.}{2012}]{Ceccarelli2012}
{Ceccarelli} L.,  {Herrera-Camus} R.,  {Lambas} D.~G.,  {Galaz} G.,
  {Padilla} N.~D.,  2012, \mnras, 426, L6

\bibitem[\protect\citeauthoryear{Ceccarelli, Padilla, Valotto \&
  Lambas}{Ceccarelli et~al.}{2006}]{Ceccarelli2006}
Ceccarelli L.,  Padilla N.~D.,  Valotto C.,    Lambas D.~G.,  2006, \mnras,
  373, 1440

\bibitem[\protect\citeauthoryear{{Ceccarelli}, {Paz}, {Lares}, {Padilla} \&
  {Garc{\'{\i}}a Lambas}}{{Ceccarelli} et~al.}{2013}]{Ceccarelli2013}
{Ceccarelli} L.,  {Paz} D.,  {Lares} M.,  {Padilla} N.,    {Garc{\'{\i}}a
  Lambas} D.,  2013, ArXiv e-prints: 1306.5798

\bibitem[\protect\citeauthoryear{{Clampitt}, {Cai} \& {Li}}{{Clampitt}
  et~al.}{2013}]{Clampitt2013}
{Clampitt} J.,  {Cai} Y.-C.,    {Li} B.,  2013, \mnras, 431, 749

\bibitem[\protect\citeauthoryear{Crocce, Pueblas \& Scoccimarro}{Crocce
  et~al.}{2006}]{crocce06}
Crocce M.,  Pueblas S.,    Scoccimarro R.,  2006, Monthly Notices of the Royal
  Astronomical Society, 373, 369{\textendash}381

\bibitem[\protect\citeauthoryear{{D'Aloisio} \& {Furlanetto}}{{D'Aloisio} \&
  {Furlanetto}}{2007}]{Daloisio2007}
{D'Aloisio} A.,  {Furlanetto} S.~R.,  2007, \mnras, 382, 860

\bibitem[\protect\citeauthoryear{{Dawson} et~al.,}{{Dawson}
  et~al.}{2013}]{Dawson2013}
{Dawson} K.~S.,  et~al., 2013, \aj, 145, 10

\bibitem[\protect\citeauthoryear{Dehnen}{Dehnen}{2001}]{dehnen01}
Dehnen W.,  2001, Monthly Notices of the Royal Astronomical Society, 324,
  273{\textendash}291

\bibitem[\protect\citeauthoryear{{Einasto}, {Einasto}, {Gramann} \&
  {Saar}}{{Einasto} et~al.}{1991}]{Einasto1991}
{Einasto} J.,  {Einasto} M.,  {Gramann} M.,    {Saar} E.,  1991, \mnras, 248,
  593

\bibitem[\protect\citeauthoryear{Ferreras \& Pasquali}{Ferreras \&
  Pasquali}{2011}]{VandeWeygaertR.2011}
Ferreras I.,  Pasquali A.,  eds, 2011, {Environment and the Formation of
  Galaxies: 30 years later}.
Astrophysics and Space Science Proceedings, Springer Berlin Heidelberg, Berlin,
  Heidelberg

\bibitem[\protect\citeauthoryear{Furlanetto \& Piran}{Furlanetto \&
  Piran}{2006}]{Furlanetto2006}
Furlanetto S.~R.,  Piran T.,  2006, \mnras, 366, 467

\bibitem[\protect\citeauthoryear{Goldberg \& Vogeley}{Goldberg \&
  Vogeley}{2004}]{Goldberg2004}
Goldberg D.~M.,  Vogeley M.~S.,  2004, \apj, 605, 1

\bibitem[\protect\citeauthoryear{Gorski, Hivon, Banday, Wandelt, Hansen,
  Reinecke \& Bartelmann}{Gorski et~al.}{2005}]{Gorski2005}
Gorski K.~M.,  Hivon E.,  Banday A.~J.,  Wandelt B.~D.,  Hansen F.~K.,
  Reinecke M.,    Bartelmann M.,  2005, \apj, 622, 759

\bibitem[\protect\citeauthoryear{Gottlober, Lokas, Klypin \& Hoffman}{Gottlober
  et~al.}{2003}]{Gottlober2003}
Gottlober S.,  Lokas E.~L.,  Klypin A.,    Hoffman Y.,  2003, \mnras, 344, 715

\bibitem[\protect\citeauthoryear{{Hamaus}, {Sutter} \& {Wandelt}}{{Hamaus}
  et~al.}{2014}]{Hamaus2014}
{Hamaus} N.,  {Sutter} P.~M.,    {Wandelt} B.~D.,  2014, ArXiv e-prints:
  1403.5499

\bibitem[\protect\citeauthoryear{{Hamaus}, {Sutter} \& Wandelt}{{Hamaus}
  et~al.}{2014}]{Hamaus2013}
{Hamaus} N.,  {Sutter} P.~M.,    Wandelt B.~D.,  2014, Physical Review Letters,
  112, 041304

\bibitem[\protect\citeauthoryear{{Higuchi}, {Oguri} \& {Hamana}}{{Higuchi}
  et~al.}{2013}]{Higuchi2013}
{Higuchi} Y.,  {Oguri} M.,    {Hamana} T.,  2013, \mnras, 432, 1021

\bibitem[\protect\citeauthoryear{{Hoyle}, {Rojas}, {Vogeley} \&
  {Brinkmann}}{{Hoyle} et~al.}{2005}]{Hoyle2005}
{Hoyle} F.,  {Rojas} R.~R.,  {Vogeley} M.~S.,    {Brinkmann} J.,  2005, \apj,
  620, 618

\bibitem[\protect\citeauthoryear{Hoyle \& Vogeley}{Hoyle \&
  Vogeley}{2004}]{Hoyle2004}
Hoyle F.,  Vogeley M.~S.,  2004, \apj, 607, 751

\bibitem[\protect\citeauthoryear{{Hoyle}, {Vogeley} \& {Pan}}{{Hoyle}
  et~al.}{2012}]{Hoyle2012}
{Hoyle} F.,  {Vogeley} M.~S.,    {Pan} D.,  2012, \mnras, 426, 3041

\bibitem[\protect\citeauthoryear{{Ilic}, {Langer} \& {Douspis}}{{Ilic}
  et~al.}{2013}]{Ilic2013}
{Ilic} S.,  {Langer} M.,    {Douspis} M.,  2013, ArXiv e-prints: 1301.5849

\bibitem[\protect\citeauthoryear{{Jennings}, {Li} \& {Hu}}{{Jennings}
  et~al.}{2013}]{Jennings2013}
{Jennings} E.,  {Li} Y.,    {Hu} W.,  2013, ArXiv e-prints: 1304.6087

\bibitem[\protect\citeauthoryear{Komatsu et~al.,}{Komatsu
  et~al.}{2011}]{Komatsu2011}
Komatsu E.,  et~al., 2011, \apjs, 192, 18

\bibitem[\protect\citeauthoryear{{Krause}, {Chang}, {Dor{\'e}} \&
  {Umetsu}}{{Krause} et~al.}{2013}]{Krause2013}
{Krause} E.,  {Chang} T.-C.,  {Dor{\'e}} O.,    {Umetsu} K.,  2013, \apjl, 762,
  L20

\bibitem[\protect\citeauthoryear{{Laureijs} et~al.,}{{Laureijs}
  et~al.}{2011}]{Laureijs2011}
{Laureijs} R.,  et~al., 2011, {Euclid Definition Study Report}, arXiv:
  1110.3193

\bibitem[\protect\citeauthoryear{{Lavaux} \& {Wandelt}}{{Lavaux} \&
  {Wandelt}}{2010}]{Lavaux2010}
{Lavaux} G.,  {Wandelt} B.~D.,  2010, \mnras, 403, 1392

\bibitem[\protect\citeauthoryear{{Lavaux} \& {Wandelt}}{{Lavaux} \&
  {Wandelt}}{2011}]{LavauxGuilhem2011}
{Lavaux} G.,  {Wandelt} B.~D.,  2011, eprint arXiv:1110.0345

\bibitem[\protect\citeauthoryear{{Little} \& {Weinberg}}{{Little} \&
  {Weinberg}}{1994}]{Little1994}
{Little} B.,  {Weinberg} D.~H.,  1994, \mnras, 267, 605

\bibitem[\protect\citeauthoryear{{Manera} et~al.,}{{Manera}
  et~al.}{2013}]{Manera2013}
{Manera} M.,  et~al., 2013, \mnras, 428, 1036

\bibitem[\protect\citeauthoryear{{Mar{\'{\i}}n} et~al.,}{{Mar{\'{\i}}n}
  et~al.}{2013}]{Marin2013}
{Mar{\'{\i}}n} F.~A.,  et~al., 2013, \mnras, 432, 2654

\bibitem[\protect\citeauthoryear{Melchior, {Sutter}, Sheldon, Krause \&
  {Wandelt}}{Melchior et~al.}{2014}]{Melchior}
Melchior P.,  {Sutter} P.~M.,  Sheldon E.,  Krause E.,    {Wandelt} B.~D.,
  2014, \mnras, 440, 2922

\bibitem[\protect\citeauthoryear{Muller, Arbabi-Bidgoli, Einasto \&
  Tucker}{Muller et~al.}{2000}]{Muller2000}
Muller V.,  Arbabi-Bidgoli S.,  Einasto J.,    Tucker D.,  2000, \mnras, 318,
  280

\bibitem[\protect\citeauthoryear{Neyrinck}{Neyrinck}{2008}]{Neyrinck2008}
Neyrinck M.~C.,  2008, \mnras, 386, 2101

\bibitem[\protect\citeauthoryear{{Padilla}, {Ceccarelli} \& {Lambas}}{{Padilla}
  et~al.}{2005}]{Padilla2005}
{Padilla} N.~D.,  {Ceccarelli} L.,    {Lambas} D.~G.,  2005, \mnras, 363, 977

\bibitem[\protect\citeauthoryear{{Pan}, {Vogeley}, {Hoyle}, {Choi} \&
  {Park}}{{Pan} et~al.}{2012}]{Pan2011}
{Pan} D.~C.,  {Vogeley} M.~S.,  {Hoyle} F.,  {Choi} Y.-Y.,    {Park} C.,  2012,
  \mnras, 421, 926

\bibitem[\protect\citeauthoryear{{Paranjape}, {Lam} \& {Sheth}}{{Paranjape}
  et~al.}{2012}]{Paranjape2012}
{Paranjape} A.,  {Lam} T.~Y.,    {Sheth} R.~K.,  2012, \mnras, 420, 1648

\bibitem[\protect\citeauthoryear{Park \& Lee}{Park \& Lee}{2007}]{Park2007}
Park D.,  Lee J.,  2007, \apj, 665, 96

\bibitem[\protect\citeauthoryear{{Peebles}}{{Peebles}}{2001}]{Peebles2001}
{Peebles} P.~J.~E.,  2001, \apj, 557, 495

\bibitem[\protect\citeauthoryear{Pisani, Lavaux, {Sutter}, {Lavaux} \&
  {Wandelt}}{Pisani et~al.}{2013}]{Pisani2013}
Pisani A.,  Lavaux G.,  {Sutter} P.~M.,  {Lavaux} G.,    {Wandelt} B.~D.,
  2013, ArXiv e-prints: 1306.3052

\bibitem[\protect\citeauthoryear{{Planck Collaboration}}{{Planck
  Collaboration}}{2013a}]{Planck2013b}
{Planck Collaboration} 2013a, ArXiv e-prints: 1303.5079

\bibitem[\protect\citeauthoryear{{Planck Collaboration}}{{Planck
  Collaboration}}{2013b}]{Planck2013}
{Planck Collaboration} 2013b, ArXiv e-prints: 1303.5076

\bibitem[\protect\citeauthoryear{Platen, van~de Weygaert \& Jones}{Platen
  et~al.}{2007}]{Platen2007}
Platen E.,  van~de Weygaert R.,    Jones B. J.~T.,  2007, \mnras, 380, 551

\bibitem[\protect\citeauthoryear{Quinn, Katz, Stadel \& Lake}{Quinn
  et~al.}{1997}]{quinn97}
Quinn T.,  Katz N.,  Stadel J.,    Lake G.,  1997, {arXiv:astro-ph/9710043}

\bibitem[\protect\citeauthoryear{{Reid} et~al.,}{{Reid}
  et~al.}{2012}]{Reid2012}
{Reid} B.~A.,  et~al., 2012, \mnras, 426, 2719

\bibitem[\protect\citeauthoryear{{Rojas}, {Vogeley}, {Hoyle} \&
  {Brinkmann}}{{Rojas} et~al.}{2004}]{Rojas2004}
{Rojas} R.~R.,  {Vogeley} M.~S.,  {Hoyle} F.,    {Brinkmann} J.,  2004, \apj,
  617, 50

\bibitem[\protect\citeauthoryear{{Rojas}, {Vogeley}, {Hoyle} \&
  {Brinkmann}}{{Rojas} et~al.}{2005}]{Rojas2005}
{Rojas} R.~R.,  {Vogeley} M.~S.,  {Hoyle} F.,    {Brinkmann} J.,  2005, \apj,
  624, 571

\bibitem[\protect\citeauthoryear{Ryden}{Ryden}{1995}]{Ryden1995}
Ryden B.~S.,  1995, \apj, 452, 25

\bibitem[\protect\citeauthoryear{{Ryden} \& {Melott}}{{Ryden} \&
  {Melott}}{1996}]{Ryden1996}
{Ryden} B.~S.,  {Melott} A.~L.,  1996, \apj, 470, 160

\bibitem[\protect\citeauthoryear{{S{\'a}nchez} et~al.,}{{S{\'a}nchez}
  et~al.}{2012}]{Sanchez2012}
{S{\'a}nchez} A.~G.,  et~al., 2012, \mnras, 425, 415

\bibitem[\protect\citeauthoryear{{Schlegel} et~al.,}{{Schlegel}
  et~al.}{2011}]{Schlegel2011}
{Schlegel} D.,  et~al., 2011, {The BigBOSS Experiment}, arXiv:1106.1706

\bibitem[\protect\citeauthoryear{{Sheth} \& {van de Weygaert}}{{Sheth} \& {van
  de Weygaert}}{2004}]{Sheth2004}
{Sheth} R.~K.,  {van de Weygaert} R.,  2004, \mnras, 350, 517

\bibitem[\protect\citeauthoryear{{Shoji} \& {Lee}}{{Shoji} \&
  {Lee}}{2012}]{Shoji}
{Shoji} M.,  {Lee} J.,  2012, ArXiv e-prints: 1203.0869

\bibitem[\protect\citeauthoryear{{Spergel} et~al.,}{{Spergel}
  et~al.}{2013}]{Spergel2013}
{Spergel} D.,  et~al., 2013, ArXiv e-prints: 1305.5422

\bibitem[\protect\citeauthoryear{{Spolyar}, {Sahl{\'e}n} \& {Silk}}{{Spolyar}
  et~al.}{2013}]{Spolyar2013}
{Spolyar} D.,  {Sahl{\'e}n} M.,    {Silk} J.,  2013, ArXiv e-prints: 1304.5239

\bibitem[\protect\citeauthoryear{{Strauss} et~al.,}{{Strauss}
  et~al.}{2002}]{Strauss2002}
{Strauss} M.~A.,  et~al., 2002, \aj, 124, 1810

\bibitem[\protect\citeauthoryear{{Sutter} et~al.,}{{Sutter}
  et~al.}{2014}]{Sutter2014c}
{Sutter} P.~M.,  et~al., 2014, ArXiv e-prints: 1406.1191

\bibitem[\protect\citeauthoryear{{Sutter}, {Lavaux}, {Hamaus}, {Wandelt},
  {Weinberg} \& {Warren}}{{Sutter} et~al.}{2014}]{Sutter2013a}
{Sutter} P.~M.,  {Lavaux} G.,  {Hamaus} N.,  {Wandelt} B.~D.,  {Weinberg}
  D.~H.,    {Warren} M.~S.,  2014, \mnras, 442, 462

\bibitem[\protect\citeauthoryear{{Sutter}, {Lavaux}, {Wandelt} \&
  {Weinberg}}{{Sutter} et~al.}{2012a}]{Sutter2012b}
{Sutter} P.~M.,  {Lavaux} G.,  {Wandelt} B.~D.,    {Weinberg} D.~H.,  2012a,
  \apj, 761, 187

\bibitem[\protect\citeauthoryear{{Sutter}, {Lavaux}, {Wandelt} \&
  {Weinberg}}{{Sutter} et~al.}{2012b}]{Sutter2012a}
{Sutter} P.~M.,  {Lavaux} G.,  {Wandelt} B.~D.,    {Weinberg} D.~H.,  2012b,
  \apj, 761, 44

\bibitem[\protect\citeauthoryear{{Sutter}, {Lavaux}, {Wandelt}, {Weinberg} \&
  {Warren}}{{Sutter} et~al.}{2014}]{Sutter2013b}
{Sutter} P.~M.,  {Lavaux} G.,  {Wandelt} B.~D.,  {Weinberg} D.~H.,    {Warren}
  M.~S.,  2014, \mnras, 438, 3177

\bibitem[\protect\citeauthoryear{{Tavasoli}, {Vasei} \& {Mohayaee}}{{Tavasoli}
  et~al.}{2013}]{Tavasoli2013}
{Tavasoli} S.,  {Vasei} K.,    {Mohayaee} R.,  2013, \aap, 553, A15

\bibitem[\protect\citeauthoryear{{Taylor}, {Vovk} \& {Neronov}}{{Taylor}
  et~al.}{2011}]{Taylor2011}
{Taylor} A.~M.,  {Vovk} I.,    {Neronov} A.,  2011, \aap, 529, A144

\bibitem[\protect\citeauthoryear{Thompson \& Vishniac}{Thompson \&
  Vishniac}{1987}]{Thompson1987}
Thompson K.~L.,  Vishniac E.~T.,  1987, \apj, 313, 517

\bibitem[\protect\citeauthoryear{Thompson \& Gregory}{Thompson \&
  Gregory}{2011}]{Thompson2011}
Thompson L.~A.,  Gregory S.~A.,  2011, ArXiv e-prints: 1109.1268

\bibitem[\protect\citeauthoryear{Tinker \& Conroy}{Tinker \&
  Conroy}{2009}]{Tinker2009}
Tinker J.~L.,  Conroy C.,  2009, \apj, 691, 633

\bibitem[\protect\citeauthoryear{{Tinker}, {Weinberg} \& {Zheng}}{{Tinker}
  et~al.}{2006}]{Tinker2006}
{Tinker} J.~L.,  {Weinberg} D.~H.,    {Zheng} Z.,  2006, \mnras, 368, 85

\bibitem[\protect\citeauthoryear{{Vogeley}, {Geller}, {Park} \&
  {Huchra}}{{Vogeley} et~al.}{1994}]{Vogeley1994}
{Vogeley} M.~S.,  {Geller} M.~J.,  {Park} C.,    {Huchra} J.~P.,  1994, \aj,
  108, 745

\bibitem[\protect\citeauthoryear{{von Benda-Beckmann} \& {Mueller}}{{von
  Benda-Beckmann} \& {Mueller}}{2007}]{vonBenda2007}
{von Benda-Beckmann} A.~M.,  {Mueller} V.,  2007, ArXiv e-prints: 0710.2783

\bibitem[\protect\citeauthoryear{{Warren}}{{Warren}}{2013}]{warren13}
{Warren} M.~S.,  2013, ArXiv e-prints

\bibitem[\protect\citeauthoryear{{Weinberg} \& {Cole}}{{Weinberg} \&
  {Cole}}{1992}]{Weinberg1992}
{Weinberg} D.~H.,  {Cole} S.,  1992, \mnras, 259, 652

\bibitem[\protect\citeauthoryear{{Weinberg}, {Mortonson}, {Eisenstein},
  {Hirata}, {Riess} \& {Rozo}}{{Weinberg} et~al.}{2013}]{Weinberg2012}
{Weinberg} D.~H.,  {Mortonson} M.~J.,  {Eisenstein} D.~J.,  {Hirata} C.,
  {Riess} A.~G.,    {Rozo} E.,  2013, ArXiv e-prints: 1201.2434

\bibitem[\protect\citeauthoryear{{Zheng}, {Coil} \& {Zehavi}}{{Zheng}
  et~al.}{2007}]{Zheng2007}
{Zheng} Z.,  {Coil} A.~L.,    {Zehavi} I.,  2007, \apj, 667, 760

\end{thebibliography}
}

\end{document}